\documentclass[sigconf, nonacm]{acmart}

\usepackage{subcaption}
\usepackage[noabbrev, nameinlink]{cleveref}
\usepackage{xspace}

\usepackage{tikz-cd} %

\usepackage{xcolor}
\usepackage{colortbl}

\definecolor{whybarcolor}{HTML}{AFDAF1} 
\newenvironment{whyquote}{
  \par\vspace{3pt}
  \noindent
  \begingroup
    \setlength{\arrayrulewidth}{2pt}%
    \arrayrulecolor{whybarcolor}
    \begin{tabular}{@{}|@{\hspace{0.7em}}p{0.95\linewidth}@{}}
}{%
    \end{tabular}%
  \endgroup
  \par
}

\definecolor{examplebarcolor}{HTML}{DCEEC5} 
\newenvironment{examplequote}{
  \par\vspace{3pt}
  \noindent
  \begingroup
    \setlength{\arrayrulewidth}{2pt}%
    \arrayrulecolor{examplebarcolor}
    \begin{tabular}{@{}|@{\hspace{0.7em}}p{0.95\linewidth}@{}}
}{%
    \end{tabular}%
  \endgroup
  \par
}

\AtBeginDocument{%
  }

\copyrightyear{2026}
\acmYear{2026}
\setcopyright{none}
\setcctype{by-nc-nd}
\acmConference[CHI '26]{Proceedings of the 2026 CHI Conference on Human Factors in Computing Systems}{April 13--17, 2026}{Barcelona, Spain}
\acmBooktitle{Proceedings of the 2026 CHI Conference on Human Factors in Computing Systems (CHI '26), April 13--17, 2026, Barcelona, Spain}
\acmDOI{10.1145/3772318.3790264}
\acmISBN{979-8-4007-2278-3/2026/04}

\begin{document}

\title[How Notations Evolve]{How Notations Evolve: A Historical Analysis with Implications for Supporting User-Defined Abstractions}

\author{Jingyue Zhang}
\affiliation{%
  \institution{Department of Computer Science and Operations Research (DIRO)}
  \institution{Université de Montréal}
  \city{Montréal}
  \state{Québec}
  \country{Canada}
}
\email{jingyue.zhang@umontreal.ca}

\author{J.D. Zamfirescu-Pereira}
\affiliation{%
  \institution{Department of Computer Science}
  \institution{UCLA}
  \city{Los Angeles}
  \state{California}
  \country{USA}
}
\email{zamfi@cs.ucla.edu}

\author{Elena L. Glassman}
\affiliation{%
  \institution{SEAS}
  \institution{Harvard University}
  \city{Cambridge}
  \state{Massachusetts}
  \country{USA}
}
\email{glassman@seas.harvard.edu}

\author{Damien Masson}
\affiliation{%
  \institution{Department of Computer Science and Operations Research (DIRO)}
  \institution{Université de Montréal}
  \city{Montréal}
  \state{Québec}
  \country{Canada}
}
\email{damien.masson@umontreal.ca}

\author{Ian Arawjo}
\affiliation{%
  \institution{Department of Computer Science and Operations Research (DIRO)}
  \institution{Université de Montréal}
  \city{Montréal}
  \state{Québec}
  \country{Canada}
}
\email{ian.arawjo@umontreal.ca}

\renewcommand{\shortauthors}{Zhang, Zamfirescu-Pereira, Glassman, Masson and Arawjo}

\begin{abstract}
Traditional human-computer interaction takes place through formally-specified systems like structured UIs and programming languages. Recent AI systems promise a new set of informal interactions with computers through natural language and other notational forms. These informal interactions can then lead to formal representations, but depend upon pre-existing formalisms known to both humans and AI. What about novel formalisms and notations? How are new abstractions created, evolved, and incrementally formalized over time---and how might new systems, in turn, be explicitly designed to support these processes? We conduct a comparative historical analysis of notation development to identify some relevant characteristics. These include three social stages of notation development: invention \& incubation, dispersion \& divergence, and institutionalization \& sanctification, as well as three functional stages: descriptive, generative, and evaluative. Within and across these stages, we detail several patterns, such as the role of linking and grounding metaphors, dimensions of meaningful variation, and analogical alignment. Finally, we offer some implications for design. 
\end{abstract}


\newcommand{\capform}[1]{\textmd{\small{#1}}}

\maketitle

\section{Introduction}

\begin{quote}
    ``The formation of abstractions `can be broken down into small, unexpected and practical sets of skills'... [S]cientific achievements, %
    [once] viewed as results of spontaneous, individual creativity... %
    become analyzable, transparent performances... %
    linked to collectively available resources and to the toolbox of signifiers in the broader culture.'' \\ \phantom{~} \hfill---Ursula Klein \cite[p.~241-2]{klein2003experiments}
\end{quote}

Human-computer interactions have historically been mediated by formally-defined structures---such as command-line interfaces, graphical user interfaces, and programming languages---that provide an unambiguous mapping to an underlying formal model. Recent advancements in AI offer a new paradigm of subjectively inferring human intent from ambiguous input~\cite{zamfirescu2025beyond}. These systems can enable rich informal interactions through natural language by leveraging existing abstractions to rapidly formalize ideas. In AI code generation, for instance, successful interactions rely on the humans' and AI models' complementary capabilities in using and translating among these abstractions. %

Yet humans collaborating together do not always defer to a pre-existing formalism. Instead, they can develop ad-hoc, new notations to ground their communication~\cite{cherubini2007let, brennan1991conversation, vaithilingam2024imagining}, treating notations more like malleable resources than rigid  systems. As current AI technologies rely upon, reproduce, and amplify established, dominant, already-formalized abstractions and notations in order to function, we wonder: %
\textit{How do humans ultimately develop new notations, new formalisms, and new abstractions, that they use to communicate with machines and each other?} 
Addressing this question could point the way to future system designs that aim to support notation creation, evolution, and incremental formalization~\cite{shipman1999incremental}.

In this work, we lay some groundwork towards addressing this question by %
presenting a parallel comparative historical analysis~\cite{skocpol1980uses} of notation development across scientific, computing, and artistic disciplines. Our analysis identifies %
33 patterns of how notations are created, evolved, and formalized over time, which are largely shared across histories and loosely categorized into three social stages of  development (invention/incubation, dispersion/divergence, and institutionalization/sanctification) and three functional stages (descriptive, generative, and evaluative). %
From our analysis, we derive a set of initial implications for the design of future systems that create new abstractions (Section~5), including that notations primarily originate through linking metaphors and most often in a social---rather than a technical---context, and that notation design decisions around what to include as ``meaningful'' (and thus what to exclude) are often left implicit by inventors, but could be made explicit and become manipulable objects through reification~\cite{beaudouin2000instrumental}.

We find that notation evolution proceeds from informal ideas that borrow from prior culture in their inventors' ready-to-hand surroundings~\cite{arawjo2020write}---especially the remixing of prior notations and the application of conceptual metaphors of linking and grounding~\cite{lakoff2000mathematics}. A notation is revised and extended as it comes into contact with diverse situations and people, and (sometimes) circulates more widely, whereupon controversies~\cite{pinch1984social} can arise and adopters may establish institutions to manage their settlement. While the usability of notation plays a mediating role in uptake, social power and suitability to environmental circumstances (material or social) also play mediating roles. Popular notations are thereafter materially and socially institutionalized, sanctifying not just the representation but the aesthetic and values encoded in their design and can serve as rhetorical boundary work~\cite{boundary-work} in a discipline. At the same time, formalized notations become raw materials that are then redeployed in new ways, forming floors in a growing tower of abstractions that together advance human knowledge. %

Our work heeds calls from HCI scholars for more historicism in the field, to better contextualize technology ``within dynamic temporal processes of emergence, change, continuity, decline, disappearance, or revival,'' and to see ``the past... as a repository of design knowledge and experience''~\cite[p.4-11]{soden2021historicism}. %
Our hope is that our investigation offers insights for future HCI research into how to support humans in defining their own abstractions and formalisms, %
rather than being limited to the designers',  creating more ``malleable'' interfaces that adapt to user needs~\cite{vaithilingam2024imagining, li2023power, xia2016objectoriented, min2025malleable}. %
We position our work in HCI and unpack terms like ``notation,'' before proceeding to our comparative historical analysis (Section~\ref{sec:history}). From there, we summarize and reflect (Sec.~\ref{sec:unified-theory}), presenting functional stages of notation development, before %
concluding with some implications and takeaways~(Secs.~\ref{sec:implications}-\ref{sec:discussion}).

\section{Context and Related Work}

\subsection{Notation in HCI and beyond}
By the term ``notation,'' we adopt the inclusive definition of Blackwell \& Green: ``a notation consists of marks (often visible, though possibly sensed by some other means) made on some medium'' \cite[p. 328]{blackwell2001cognitive}. Written languages on paper, programming languages on screens, algebraic equations, Venn diagrams, finite automata diagrams, tables, bar charts, chemical formulas, JSON---all of these are considered notations for our present purposes. Although a notation could comprise a few isolated marks without rules about interactions between them, here we are primarily focused on written notational \textit{systems} similar to the manner of Goodman~\cite{goodman1976languages}: systems of marks with syntax and semantics that (try to) delineate ``valid'' members of the notation such that each have an interpretation. %

The use of notation happens everyday in small ways, e.g., whenever people work together over a whiteboard or paper towards a joint objective. People jot down X's, boxes and arrows to stand-for concepts they are working through. As people find some notations useful to their thinking or coordination, people begin to `formalize' the notation---making explicit the rules of the system, to disambiguate meaning and anticipate misinterpretations~\cite{cherubini2007let}. The created notation can then outlast the original discussion, become embedded into computer systems and even inspire new inquiries, making some ideas easier to imagine than others \cite{blackwell2001cognitive, iverson2007notation, arawjo2020write, klein2003experiments}. Almost everything we do with computers involves notations.

Notations are deployed and embedded throughout the process of HCI and software development. For instance, developers use UML diagrams %
~\cite{platt2015evolution}, flow charts~\cite{arawjo2020write}, programming languages, and ``double-diamond'' design models%
~\cite{designcouncil_double_diamond_history}.  %
The practice of interface design broadly requires developing notations for the user interface~\cite{blackwell2001cognitive, saquib2021embodiedmath, xing2016energybrushes}, especially when designing computational media authoring software. %
EuterPen~\cite{cavez2025euterpen} allows composers to %
shift notes between typeset and handwritten forms; the interface reifies formality by making handwritten notes easier to remove. Some systems work studies how new developments in pen-based hardware and deep learning models might reconfigure programming practices~\cite{codeshaping, arawjo2020write}. %
Notations, as interfaces, also afford certain actions and resist others~\cite{blackwell2001cognitive}. Visual data-flow environments like Blender's support readability and chaining of operations, but resist operations that require iterative loops. %

Studies are also conducted on various existing notating practices, usually in specific domains (e.g., how programmers draw diagrams to communicate ideas~\cite{hayatpur2024ascii, hayatpur2025shapesdiagrams, cherubini2007let}). Studying software teams, Cherubini et al.~\cite{cherubini2007let} found a ``tendency to adopt informal, ad-hoc notations'' and a ``limited adherence to standards of any sort.'' They show how, as conversations expand from internal, ad-hoc dialogues over a whiteboard to communicating with clients and upper managers, notations are similarly systematized, from ad-hoc, informal, transient sketches to ``reiterated'' sketches that can become a ``rendered drawing'' when a sketch becomes ``so important that it warranted the investment of time and effort to transform it to a more permanent form'' \cite[563-4]{cherubini2007let}. Researchers have done rich historical investigations of \textit{individual} notations (e.g., \cite{kaiser2019drawing, klein2003experiments, arawjo2020write, platt2015evolution, klabnik2016history}), but the more \textit{general} mechanisms and patterns through which new notations are created and formalized are less understood. %

Many notations are culturally learned and inherited. When we in HCI speak of ``formalizing'' a design idea, we tend to implicitly frame it in terms of translating informal ideas into \textit{existing} ``formal'' representations: from natural language to a programming language, for instance~\cite{shipman1999incremental, shipman1999formality, maes_shneiderman_debate}. This meaning is what Shipman and McCall largely deploy in their papers on incremental formalization~\cite{shipman1999incremental, shipman1999formality}. %
``Notation'' as a term is also typically understood in common usage as an already-formalized system of inscription: math notation, molecular formulae, and the like. 
Yet what seem today as obvious notations often have relatively short histories: for instance, arrows in diagrams emerged around the 18th century. %
Seemingly ``obvious'' notations to academics are also not obvious to everyone: e.g., about one-third of the U.S.~and German populations have low literacy in reading data visualizations~\cite{galesic2011graph}. %
Even when people cannot read them, formalized notations often serve a rhetorical purpose---``an impression of rationality''~\cite[p. 1]{lynch1991pictures} and \textit{ethos} conveying credibility and authority.

To understand notations even more fully, we attend to how they are \textit{socially constructed} beyond their functional or technical character. Semiotics and cognitive science will provide us tools to unpack notation and what we mean by ``formal'' and ``formalization.''

\subsection{Semiotic, cognitive, and social perspectives on notation and formality} %

From the vantage point of semiotics and science \& technology studies (STS), a notation is an abstraction with a representational system that models some empirical, observable phenomena. The notation brings the phenomenon into the ``lab,'' where it can be picked apart, manipulated, read, presented, compared, and combined~\cite{latour2011drawing}. Good notational systems produce theories through their usage that map predictably to states of the world. For instance, in science, notations suggest experiments that can be empirically verified; in music, notations coordinate the (dimensions of) reproducibility of performances. %
As such, it is fruitful to adopt a triadic framing from classical semiotics\footnote{A full discussion of semiotics is out of scope here, but useful references are Ogden \& Richard's triangle of reference~\cite{ogden1927meaning} and Pierce's sign/object/interpretant triad~\cite{zeman1977peirce}.} that separates the parts of sign, thought, and referent that ground notational systems embedded in social life:

\begin{figure}[h]
    \centering
    \includegraphics[width=0.7\linewidth]{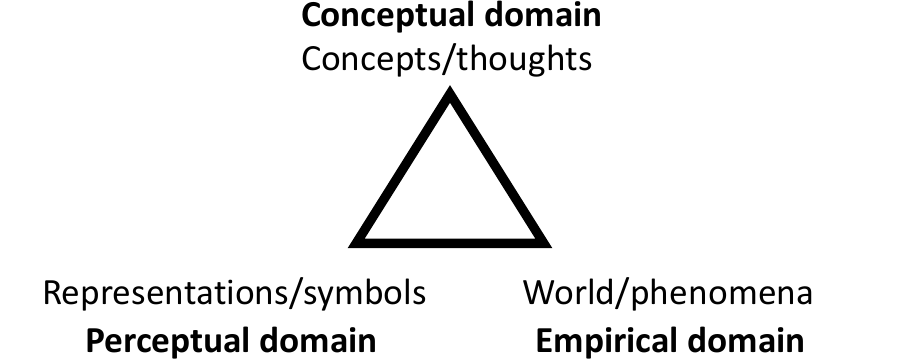}
    \label{fig:semiotic-triad}
    \Description{Triadic diagram illustrating a semiotic framework for notation. A triangle connects three domains: at the top, the Conceptual domain labeled “Concepts / thoughts”; at the bottom left, the Perceptual domain labeled “Representations / symbols”; and at the bottom right, the Empirical domain labeled “World / phenomena.” The diagram emphasizes the relationships between concepts, representations, and real-world phenomena in grounding notational systems.}
\end{figure}

From experience in the world (phenomena W), we form concepts and theories (C). Commonplace concepts are represented (R) with oral or written language. Systems of complex phenomena can inspire writing systems that help us cognitively manage the system's complexity, remember situations, and calculate or anticipate the result of physical processes. These ``tools of thought'' \cite{iverson2007notation} or ``paper tools'' \cite{klein2003experiments} can inspire new areas of inquiry, prompting the mobilization and assembling of entities in the physical world to ``perform'' or ``test'' the ``theory'' created by a given arrangement of lines on the page. Thus, what is at first a movement from the empirical realm to representation (W$\rightarrow$R) can act back upon the world (R$\rightarrow$W); similarly, representations can, through reflection, inspire new concepts through abstraction, such as grouping ``similar-looking'' notations into a common class (R$\rightarrow$C). %
Though a simplistic abstraction, separating W, C, and R can help us clarify how developments in one plane---say, conceptual understanding of world phenomena---can affect another. %

From the perspective of cognitive science, notations are a kind of embodied metaphor to abstract phenomena. Lakoff \& N\'{u}\~{n}ez %
argue that mathematical notation emerges from repeated application of \textit{conceptual metaphor}: ``%
a neural mechanism that allows us to use the inferential structure of one conceptual domain (say, geometry) to reason about another (say, arithmetic)'' \cite[p. 6]{lakoff2000mathematics}. %
Metaphors in mathematics are initially \textit{grounding metaphors}, in that they ground abstractions in everyday (graspable) human experience, such as notions of counting, quantity, and inside/outside relationships that babies learn around one year of age. %
Once created, abstractions can then be later used to inspire new abstractions---a \textit{linking} metaphor---analogically building abstractions layer by layer, forming the system of mathematics we know today. For instance, Boole's set theory derives from attempts to analogically align sets with laws of arithmetic such as associativity and commutativity---quite literally to imagine, ``what if sets are arithmetic?''---and then cognitively go through the implications, amending the notational system from there~\cite[p. 142]{lakoff2000mathematics}. Although Lakoff \& N\'{u}\~{n}ez only examine mathematical notation, their notions of conceptual metaphor will prove useful to understanding notations more broadly.

\begin{figure}
    \centering
    \[
\begin{tikzcd}
I \arrow[r, "t"] & F \\
 & I \arrow[u, "c"]
\end{tikzcd}
\]
    \caption{The `square of formality,' revealing a hidden social dimension to formalization. Informal ($I$) becomes formal ($F$) through `horizontal' (translation $t$ to existing form) and `vertical' (creation $c$ of new form) movements, representing two types of formalization processes. In practice, movements across the square of formality often mix (adoption of existing notations, mixed with new ones).}
    \label{fig:square-of-formality}
    \Description{ "The Square of Formality": A simple 2x2 grid diagram showing the relationship between informal and formal states through different types of transitions. The diagram shows four quadrants arranged in a square: Top left: "I" (Informal), Top right: "F" (Formal), Bottom left: "I" (Informal), Bottom right: "F" (Formal). Two types of arrows connect these quadrants: Horizontal arrow "t" (translation): Shows movement from informal to formal through translation to existing established forms. Vertical arrow "c" (creation): Shows movement from informal to formal through creation of new forms that gain power and credibility through social processes}
\end{figure}
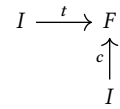

Finally, useful for our analysis is a critical reckoning with the term ``formality,'' which is almost always deployed in computing fields without further explanation. We perform an etymological analysis in Appendix~\ref{appendix-formality}, revealing social dimensions to formality, such as early connotations to law and ceremony, and later connotations of the loss of meaning. Ultimately, formalization is accomplished through two primary ``movements'' that are not necessarily mutually exclusive: \textit{translation} to existing, established forms, and \textit{creation} of new forms that gain power and credibility through social processes (Figure~\ref{fig:square-of-formality}).

\subsection{Related Work on Usability, Customization, and Power}
Our work picks up a line of early research from Green, Petre, Blackwell and others~\cite{blackwell2001cognitive} on the usability of interfaces as cognitive technologies. Their Cognitive Dimensions of Notations~\cite{blackwell2001cognitive} framework provides a set of %
usability heuristics to analyze existing and proposed graphical \textit{interfaces}, which are defined as much by their `environments' and `mediums' as their particular pre-defined notations~\cite{cdtutorial}. 
Our later observations about mappings between the domain and (representational) notations adds to the Cognitive Dimensions’ discussion of the `closeness of mapping.’
Here, we focus on the \textit{process} of notation development---their historical creation, evolution, and collective refinement.

Recent work in GUI interface customization has given users unprecedented flexibility without resorting to writing code, but it still falls short of supporting the development of entirely new notations. For example, DynaVis~\cite{vaithilingam2024dynavis} dynamically generates custom GUI widgets from pre-defined elements based on users’ natural language commands and Meridian~\cite{meridian} supports the design and implementation of malleable overview-detail interfaces where users can customize how attributes show up across views. With technology like LLMs, malleable interfaces can even generate and incorporate new attributes into views in response to user queries~\cite{min2025malleable}, but the interaction by which these attributes are defined and computed over is still constrained to existing natural and programming languages.

More recently, Li et al.'s analysis of creativity support tools~\cite{li2023beyond} included empirical evidence that systems' predefined abstractions and notations are reifying an on-going power imbalance between system designers and users---regardless of how carefully those abstractions and notations are designed. 
The process by which people design their own languages for expression, which we examine in this paper, provides examples of this process that can potentially guide future system designers in ways that might mitigate this power imbalance by supporting not just user-defined abstractions but user-defined notations as well.

\section{Stages and Patterns of Notation Development} \label{sec:history}

To better understand the process of formalization of new notational systems, we conducted a comparative historical analysis of the development of different notations which individually have been documented in prior literature. Specifically, we conduct a \textit{parallel comparative history} which ``seek[s] above all to demonstrate that a theory similarly holds good from case to case... [and where] differences among the cases are primarily contextual particularities against which to highlight the generality of the [theorized] processes''~\cite[p. 178]{skocpol1980uses}. Our methodology to seek out ``broad coverage in [the] selection of cases''~\cite[p. 179]{skocpol1980uses} and iteratively comparing and contrasting these cases to emerge at general patterns follows \textit{inductive iteration} in comparative-historical research~\cite{yom2015methodology}, with its ``cyclical logic of moving back and forth between data and theory, as researchers craft the most plausible deterministic explanation to account for observed outcomes in all cases''~\cite[p. 618]{yom2015methodology}. %
This is an interpretivist methodology~\cite{crabtree2025h} common to historical, humanities, and qualitative work, and shares camaraderie with the practice of ``using historical case studies'' to arrive at theory in science \& technology studies~\cite[p. 6]{soden2021historicism}, such as the comparative analysis of bicycle histories by Pinch and Bijker~\cite{pinch1984social}.%

Since histories of specific notations tends to miss detailed, direct observations around the initial creation process, we complement this ``macro'' analysis with occasional references to experiment-based literature from experimental semiotics, communication theory, and cognitive science into how people use notations to ground communication, largely in lab studies. %
We focus here on notations that %
have a well-established character and purpose. %
The following questions guided our review:

\begin{enumerate}
    \item Inciting Factors: Why does a notation emerge? What compels or necessitates the creation of a new notation?
    \item Creation Process: How is a notation initially created? Are there common steps or techniques?
    \item Evolution: What stages does a notation go through until it is considered sufficiently ``formalized''?
    \item Purpose of Formalization: Why and when is formalization helpful, and when is it not?
\end{enumerate}

We began with a diverse set of notations across five disciplines---music, dance, chemistry, physics, and computer programming---that had prior historical literature to draw upon in our initial analysis.\footnote{Computing systems encode and surface many notations, and users come from all walks of life, not just technical backgrounds---thus notations co-invented with future systems could be for a range of purposes. As such, we did not want to artificially restrict ourselves to notations just adjacent to HCI.} As our work continued and we discussed with others, we gradually expanded to other notational domains, such as sign language writing systems, UML diagrams, Markdown, and juggling notation, in an iterative manner to reach saturation on themes and discover any themes we had missed.\footnote{Saturation is judged to be achieved when no or little new patterns emerge from viewing more cases~\cite{charmaz2017constructivist}. For instance, our late investigation of quantum circuit notation, conducted towards the end of our inquiry, merely confirmed existing patterns of linking and grounding metaphors, cultural context informing design, and routinization.} %
Appendix~\ref{appendix-notations} and Table~\ref{tab:notation-survey} list all notations considered. We first relied upon previous, detailed historical analyses of specific notation development (e.g.,~\cite{kaiser2019drawing, klein2003experiments}), %
and from these histories looked at primary sources to clarify original representations, such as Lewis's structural formula~\cite{lewis1916atom} and Booch's early notation for software design~\cite{booch1981describing}. 

Our review identified many empirical \textit{patterns} in the notation development process. %
We state each pattern, briefly describe it, and provide examples. A few patterns require more lengthy unpacking; in these cases, we connect to related work. These patterns are not exhaustive---like how the CDs framework was refined over decades~\cite{blackwell2001cognitive}, our work is an initial step, rather than a conclusive theory. To structure our patterns and aid presentation, we assign patterns to the three overarching social \textit{stages} of notation development. While purposes and usage of notations vary widely, stages are remarkably similar across historical trajectories. The stages of \textit{social} development of a notational system appear as follows: 

\begin{itemize}
    \item Stage 1: Invention and incubation
    \item Stage 2: Dispersion and divergence
    \item Stage 3: Institutionalization and sanctification
\end{itemize}

\begin{figure*}
    \centering
    \includegraphics[width=0.95\linewidth]{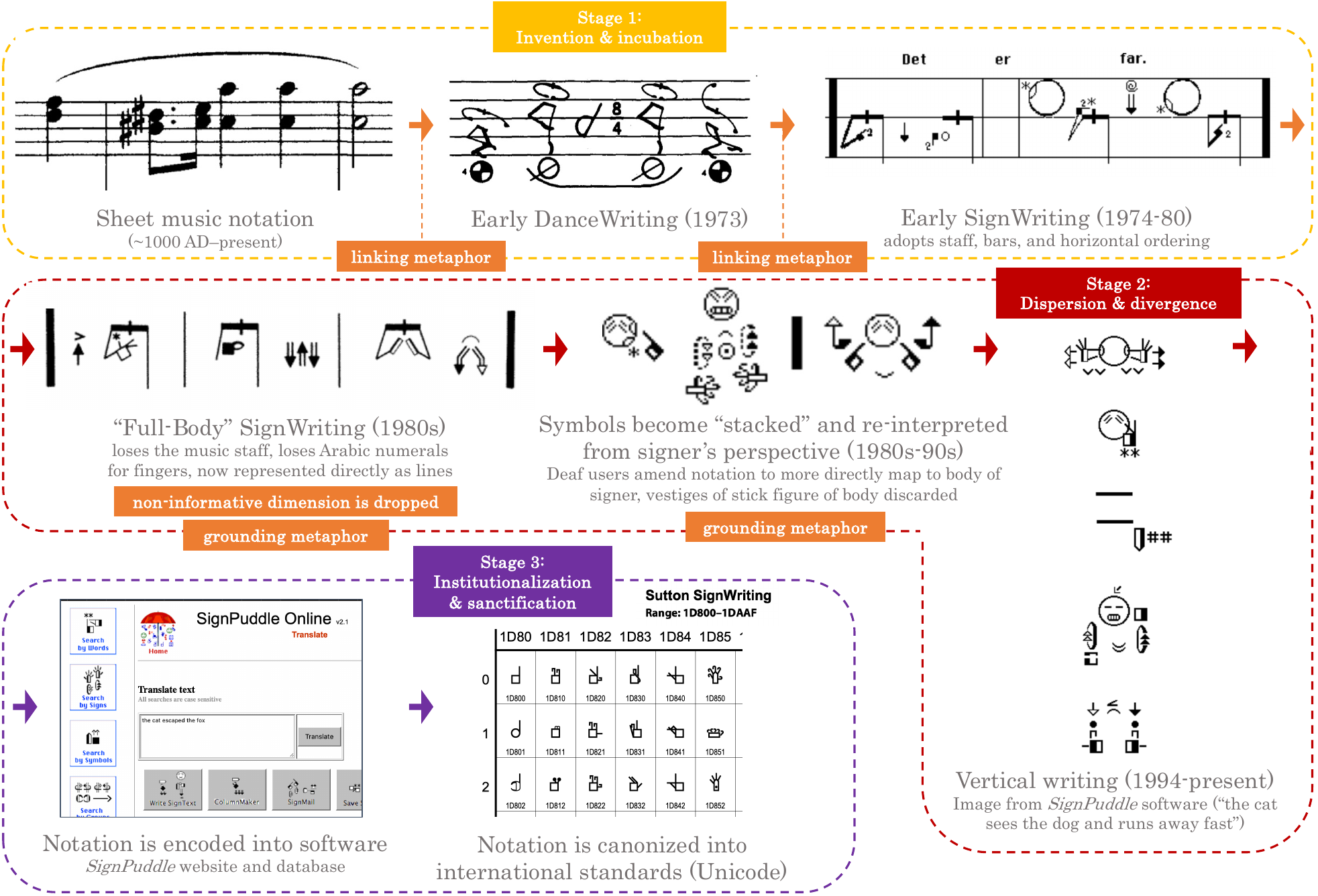}
    \caption{The evolution of \textit{SignWriting}, a notation for writing sign languages originally invented by Sutton and co-designed with Deaf communities and academic researchers over decades. Tracing its lineage back to sheet music notation, we see the prominent use of linking metaphors and grounding metaphors~\cite{lakoff2000mathematics} throughout, including new metaphors that break fundamental aspects of the previous: music staffs gradually fade away, interpretation is drastically changed without affecting iconography, and writing shifts from horizontal rows to vertical columns. SignWriting is materially and socially institutionalized in software, educational resources, and encoded into Unicode standards. Changes now demand costly reprints, re-encodings, and mobilization of committees. Images reproduced from Sutton~\cite{sutton2015history, sutton1973sutton} under fair use for purposes of academic analysis.} 
    \label{fig:signwriting}
    \Description{ A flowchart showing the historical evolution of SignWriting notation across three developmental stages, with arrows connecting different notation systems and examples of their visual representations. Left column (Stage 1: Invention & Incubation): Shows sheet music notation from around 1000 AD with traditional musical staff lines, notes, and bars. An arrow labeled "linking metaphor" points to Early DanceWriting (1973), which adapts the musical staff system with stick figures overlaid on horizontal lines. Middle column (Stage 2: Dispersion & Divergence): Shows the progression from Early DanceWriting to Early SignWriting (1974-80), where the musical staff structure is adapted for sign language with symbols representing hand positions. The notation then evolves to "Full-Body" SignWriting (1980s), losing Arabic numerals for fingers and representing them directly as lines. Another evolution shows symbols becoming "stacked" and re-interpreted from the signer's perspective (1980s-90s), with Deaf users amending the notation to map more directly to the body of the signer. Right column (Stage 3: Institutionalization & Sanctification): Shows Vertical Writing (1994-present) where the notation shifts from horizontal rows to vertical columns. The final example shows an encoded image from SignPuddle software displaying "the cat sees the dog and runs away fast" in vertical SignWriting notation. The notation is canonized into international standards (Unicode). The figure demonstrates key concepts like grounding metaphors, linking metaphors, and how non-informative dimensions are dropped over time as the notation evolves and becomes institutionalized.}
\end{figure*}

These stages form a spectrum and are not rigid boundaries. We clustered patterns into the most relevant stage for ease of presentation; however, patterns can be applicable across stages. Figure~\ref{fig:signwriting} illustrates the evolution of one notation, Sutton's SignWriting, tracing its lineage to sheet music notation and connecting to patterns in the main text. 

To help readers navigate the information, we additionally clustered these patterns and arrived at four primary categories:

\definecolor{cognitive}{RGB}{177, 218, 239}    %
\definecolor{social}{RGB}{255, 236, 187}       %
\definecolor{formal}{RGB}{221, 238, 200} %
\definecolor{constraint}{RGB}{220, 196, 221}    %

\newcommand{\fixheight}[1]{\rule[-0.3ex]{0pt}{2.2ex}}

\newcommand{\cognitive}[0]{\colorbox{cognitive}{Cognitive\fixheight{}}}
\newcommand{\social}[0]{\colorbox{social}{Social\fixheight{}}}
\newcommand{\formal}[0]{\colorbox{formal}{Formal\fixheight{}}}
\newcommand{\constraint}[0]{\colorbox{constraint}{Constraint\fixheight{}}}

\begin{enumerate}
    \item \textit{\cognitive{}}: Cognitive factors around a notation’s inspiration, design, and evolution.
    \item \textit{\social{} Credibility and Valuation}: Social factors around the notation’s adoption, spread, and maintenance, and how it is valued or confers value. 
    \item \textit{\formal{}ization Processes}: How a notation is refined, disambiguated, and extended, and processes related to it.
    \item \textit{\constraint{}}: Material or cultural factors that constrain a notation's direction of evolution. 
\end{enumerate}

\noindent We add colored badges to aid readability. Like stages, these categories aren’t rigidly defined but are included as visual aids, e.g., cognitive factors certainly also have cultural dimensions. %

Cognitive patterns will require us to unpack relevant concepts from embodied mathematics~\cite{lakoff2000mathematics}, Variation Theory~\cite{marton1981phenomenography, bussey2013variation, marton2014necessary}, and Structure-Mapping Theory in cognitive science~\cite{gentner2017analogy}: the prominent roles of conceptual metaphors of linking and grounding, analogical alignment, and how notations surface and map dimensions of meaningful variation in the conceptual/world domain to the dimensions of meaningful \textit{perceptual} variation in the representational domain. Throughout this section, we present snapshots from the original copyrighted material under fair use for the purpose of academic commentary; in each case, we cite the original source. %

\subsection{Invention and incubation (stage 1)}

\subsubsection{\cognitive{} \social{} \bf Notations are invented to manage complexity and coordinate the reproducibility of action}\label{sec:manage-complexity}  What kind of situations compel the creation of a new notation? The most prominent reasons were: to make complexity manageable, %
to store information, %
to ease comparison, to communicate efficiently, to remember terms during mental calculation, to reduce writing effort, and to coordinate reproducible actions across space and time. These underscore notation's usage as a mnemonic device, storage device, comparison tool, calculating tool, shorthand, coordination device, and framing device.

For example, Venn invented his eponymous diagrams as a book-keeping device %
for ``any one who wishes to grapple effectively with'' the combinatorial explosion of mathematical terms in logic expressions~\cite[p. 7]{venn1880diagrammatic}. Feynman similarly invented his diagrams ``as a bookkeeping device for simplifying lengthy calculations'' in quantum electrodynamics~\cite[p. 4]{kaiser2019drawing}.
Other notations like music and dance aim to store precise actions across space and time, such that later generations can produce structurally similar performances. Sutton invented DanceWriting,  concerned that an ``old, specialized form of [Danish] ballet training... could be lost or forgotten,'' %
emphasizing efficiency, reduction of effort, readability and learnability for non-artistic users~\cite[p. 1-2]{sutton1973sutton}. Finally, Playfair invented line charting so that %
``as much information may be obtained in five minutes as would require whole days to imprint on the memory... %
by a table of figures'' \cite[p. xii]{playfair1786commercial}.

\subsubsection{\social{} \bf The same problem domain can prompt independent development of notations to address it}\label{pat:independent_development} Since a notation emerges to manage complexity and coordinate action, 
recurring needs to manage complexity (as well as scientific and technological advancements) can prompt the independent, sometimes contemporaneous invention of notations for the same task.  %
Examples include Newton and Leibniz's calculus notations; Feynman-Dyson diagrams and contemporaneous Koba \& Takeda’s ``transition diagrams'' developed in Japan around the same time for the same purposes~\cite{kaiser2019drawing}; and Dalton's atom diagrams and Berzelius' chemical formulas~\cite[p. 40]{klein2003experiments}. In software engineering, ``by the late 1980s there were more than fifty separate modeling languages---each with their own syntax, structure and notation,'' leading to ``alarmingly high project failure rates,''~\cite[p. 348-9]{platt2015evolution} which prompted an effort to ``unify'' the languages (UML).

\subsubsection{\social{} \constraint{} \bf Inventors adapt commonplace representations in their ready-to-hand surroundings and background}\label{pat:ready_to_hand} %
Inventors are influenced by their cultural, disciplinary, and personal background, and adapt representations and aesthetics already familiar to them. %
For instance, Arawjo recounted that the flow chart method of Goldstine \& von Neumann was inspired by the electrical schematic block diagrams common to the Moore School of Electrical Engineering, where the project took place~ %
\cite{arawjo2020write}. In physics, Penrose diagrams were not just adapted from Minkowski spacetime diagrams, but inspired by the art of M.C. Escher~\cite{wright2013origins}. The resemblance is not just visual---the \textit{concepts} were easy to align, as well. Escher's drawings show ``impossible objects'' and distort perspective to render infinity, while Penrose's diagrams of conformal transformations of spacetime in general relativity similarly ``so distorted the relationships between distances on the page that `infinity' could be represented as finite''~\cite[p. 137]{wright2013origins}.\footnote{According to Structure-Mapping Theory, if the resemblance does not correspond to alignable concepts, this cross-mapping would interfer with rather than assist in cognition with the new notation~\cite{gentner97}.} %

\subsubsection{\cognitive{} \bf Notations foreground the dimensions of variation in the problem domain deemed most meaningful to inventors, making it easier to read and encode details that best align these dimensions, and harder or impossible to capture details that fall outside them.}\label{pat:dimensions_of_variation} %
Salient features of a notation map to (what the inventors believe is) the most important features of the modelled domain; left out or implied elements are less critical or deemed irrelevant. When critical dimensions of variation are left out, the notation must be extended or reconceptualized. 
Variation Theory~\cite{bussey2013variation} suggests that, if included, irrelevant dimensions of variation could distract users by capturing coincidental similarities and differences between concepts being represented, or mislead users with their consistency into thinking that the particular fixed value along that dimension is critical. %

Invariant dimensions are typically abstracted or omitted over time. For instance, in Sutton's SignWriting, the position of body parts are critical sites of variance, represented with stick figures. However, Deaf adults using SignWriting ``[felt] strongly that [a] stick figure is not necessary,'' reacting against a ``full body'' version of SignWriting that always included a circle for a head regardless of whether facial expressions held meaning~\cite[p. 42]{sutton2015history}. The shoulder and chest lines were eliminated, with the head only displayed if relevant to the sign's meaning (Figure~\ref{fig:signwriting}).\footnote{Interestingly, hearing parents of Deaf children preferred the variant for learning purposes, perhaps because it makes the mapping to the body more explicit, a mapping that is tacit and thus unnecessary for experienced signers. Structure-mapping theory shows that users' prior knowledge informs %
which distinct signs are compared and distinguished~%
~\cite{gentner97}.} This aligns with Clark \& Brennan's ``principle of least  effort'' in communicative grounding: ``don't expend any more effort than you need to get your addressees to understand you with as little effort [as possible]'' \cite[p. 226]{clark1991grounding}. By removing the need to write invariant (or typically invariant) features, one reduces physical effort; this may be one mechanism through which iconic representations become more abstract~\cite{fan2023drawing}.

A dimension of variation may be encoded at different levels of fidelity. 
Consider Labanotation: The ``degree'' of a movement is specified by three shape fill-in patterns (unfilled, filled, or striped) corresponding to positions on floor, parallel to floor, and up high~\cite{knust1959introduction}:

\begin{center}
\includegraphics[width=0.4\linewidth]{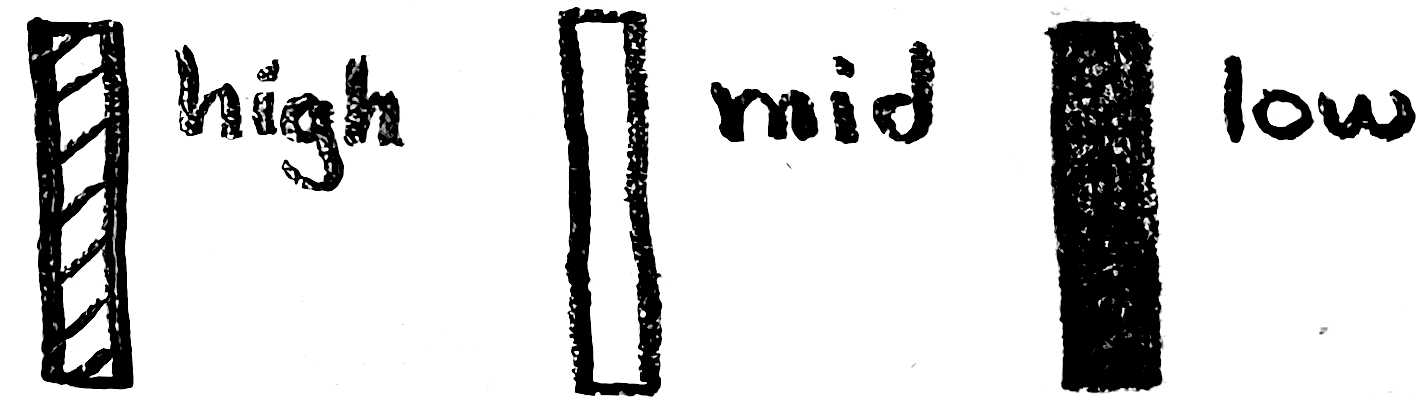}
\end{center}

\noindent These discrete directions are easy to grasp at a glance, but cannot specify precise angles such as "15 degrees." From the success of the notation, we can infer that, for most purposes in European choreography in the 20th century, this imprecision did not pose a problem. However, if one were to create a score that required precise angles---say, to align dancer's legs with ticks of a clock---Labanotation would struggle to represent the movement.

\subsubsection{\cognitive{} \bf Notations leverage innate and cultural perceptual channels, saturating them with additional meaning.} \label{perceptual-channels} A notation maps meaning to perceptual channels. These channels are \textit{innate cognitive} or \textit{learned cultural} perceptual schemata. Innate channels are biological affordances of human visuospatial cognition, which include \textit{entity-level} aspects like color, shape, size, orientation, and texture, and \textit{relational} (gestalt) aspects like ``touching,'' connection, proximity, arrangement, quantity, order, overlay, containment (concepts of inside/outside), and trajectory~\cite{lakoff2000mathematics}. Examples of learned cultural channels are the Latin alphabet and Arabic numerals, which form bounded orders and sets of symbols. When used in a notation, channels often take on a digital (rather than analog) semantics---either a symbol is inside a circle or not; either entities touch or not; either the $x$ is on the left-hand or right-hand side of the $=$. This follows Goodman's insight that notational systems rely upon digital interpretations enforced by (tacit or explicit) boundaries between symbols~\cite{goodman1976languages}---e.g., we see ``o'' and a ``$\sigma$'' not as points on a spectrum but as separate entities. The notion of perceptual channel connects with taxonomies of marks and visual channels in data visualization~\cite{munzner2014visualization}. Patterns in Stage~2 cover how channels are leveraged and extended during notation development.

\subsubsection{\cognitive{} \bf Grounding metaphors} \label{grounding-metaphors}
Embodied perceptual experience plays a critical role in notation design as well as judging its usability. \textit{Grounding metaphors}~\cite{lakoff2000mathematics} can be separated into types, from \textit{spatial metaphors}, such as how microgesture notation maps to and visually resembles the position of fingers on hands~\cite{micro-gesture-notation-2023}, to \textit{embodied metaphors}, i.e., perceptual affordances~\cite{gaver1991technology} that invite manipulation. For spatial ``closeness of mapping'' (Cognitive Dimensions of Notation~\cite{blackwell2001cognitive}), consider how microgesture notation is constructed to resemble how the thumb is set apart from the other fingers---the leftmost digit is `lower' \cite{micro-gesture-notation-2023}: 

\begin{center}
\includegraphics[width=0.5\linewidth]{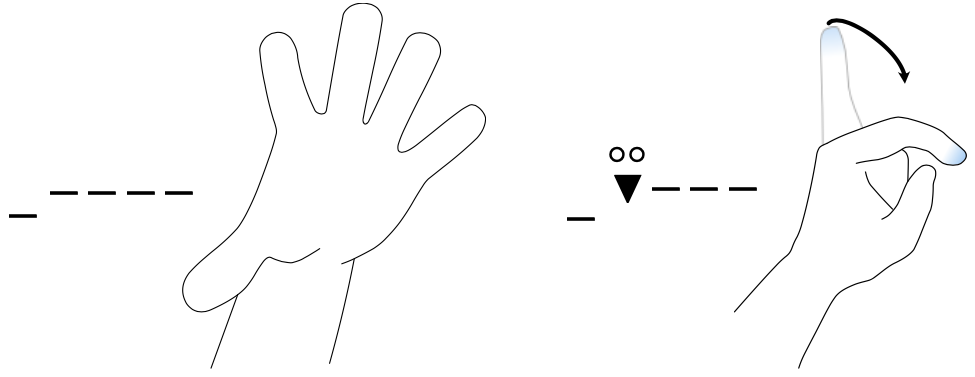}
\end{center}

\noindent For embodied metaphors, consider the quantum computing notation of Coecke and Kissinger~\cite{coecke2017picturing}, who extended a previous string diagram notation which represents functions as boxes on wires. The inventors cut off a corner of the function-box to enable perception of shape rotation so that now, ``we can slide boxes around on wires like beads on a necklace''~\cite[p. 109]{coecke2017picturing}: %

\begin{center}
\includegraphics[width=0.4\linewidth]{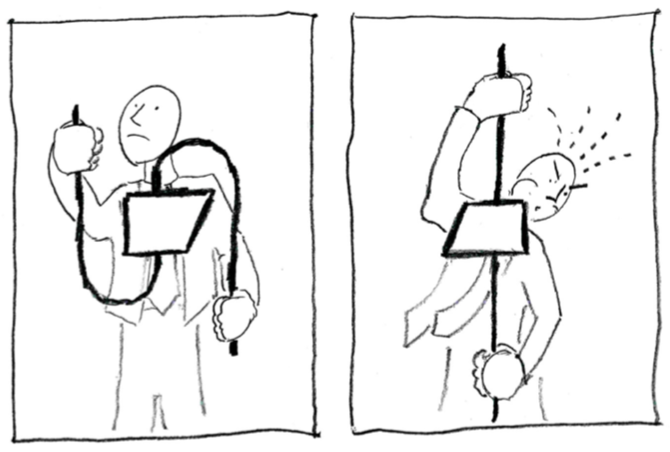}
\end{center}

\noindent The visual resemblance of the notation to beads on strings invites sliding and yanking; its creators take advantage of this to assign meaning to shape rotation as taking the transpose of a function~\cite[p. 109]{coecke2017picturing}.%

In McColloch \& Pitt's perceptron work that forms the foundation of neural networks, authors made diagrams that resemble neurons under the microscope~\cite{mcculloch1943logical} in the process of deriving a logical calculus of brain activity. Kleene later adopted their ``net'' notation to prove formal automata equivalent to regular expressions  \cite{kleene1956representation}. %

Feynman diagrams present a contrary case where a visual resemblance can lead users astray---in terms of interface design, a kind of grounding \textit{metaphor mismatch}~\cite{neale1997role}, also called a cross-mapping~\cite{gentner97}. Feynman diagrams emerged from a linking metaphor to Minkowski spacetime diagrams, which initially went over poorly: ``Bohr raised repeated objections to the very notion that spacetime diagrams could be of any help for studying... %
inherently unvisualizable quantum phenomena''%
~\cite[p. 47]{kaiser2019drawing}.\footnote{Physicists worried the diagram's visual similarity to discrete particles would mislead~%
\cite{kaiser2019drawing}, an interpretation that Feynman %
disavowed~\cite[p. 6]{wuthrich2010genesis}. Yet curiously, Dyson makes the grounding metaphor when he introduced the diagrams to physics: ``In Feynman's theory the graph corresponding to a particular matrix element is regarded, not merely as an aid to calculation, but as \textit{a picture of the physical process} which gives rise to that matrix element''~\cite[p. 496; emph. added]{dyson1949radiation}.} Today, the notation's resemblance to particle scattering tracks %
keeps %
``working physicists and teachers... busy trying to prevent colleagues and students from making incorrect interpretations on the basis of that similarity''~\cite[p. 10]{wuthrich2010genesis}.

\subsubsection{\cognitive{} \bf Linking metaphors} \label{linking-metaphors}
The second major type of conceptual metaphor~\cite{lakoff2000mathematics} is a \textit{linking metaphor}, whereby a new abstraction emerges from a metaphor to an existing one. Linking metaphors are extremely common during the invention of new notation. Examples of linking metaphors include molecular formula from algebraic notation, dance notations from music notation, Feynman diagrams from Minkowski spacetime diagrams, and flow charting for the ENIAC computer from block diagrams in electrical schematics~%
\cite{arawjo2020write}. Conceptual properties of one system, not just its representation, are applied to the new task domain, generating candidate inferences~\cite{gentner2017analogy}. %

Linking metaphors can help users, familiar in the source notation(s), to learn the new one. For example, consider Playfair's introduction of so-called ``charts'' for data visualization---the term itself originally meant a \textit{map} for navigation~\cite{oxfordenglishdictionary}. Playfair's charts abstracted the \textit{distance} metric of geographic maps: now north-south and east-west stand not for distance, but for years and wealth. Playfair %
explicitly mentions his inspirations: %
``To those who have studied geography, or any branch of mathematics, these charts will be perfectly intelligible''~\cite[ix]{playfair1786commercial}. Playfair notes that a French nobleman presented his book to King Louis XVI ``%
who, being well acquainted with the study of geography, understood it readily, and expressed great satisfaction''---a connection which then, Playfair contends, helped him gain legal privileges for spreading his work in France. 

Linking metaphors can be chained, i.e., notations can build on one another (Figure~\ref{fig:signwriting}). Sutton's SignWriting ``evolved from DanceWriting''~\cite{sutton2015history}, a dance notation by Sutton that was inspired by an earlier dance notation by Zorn~\cite{signwriting2014} that made a linking metaphor to sheet music notation, superimposing stick figures on a 5-line staff notation. %

\subsubsection{\cognitive{} \bf To leverage new metaphors, notation may need to break alignment with past metaphors.}\label{pat:break_alignment} %

Both linking and grounding metaphors rely upon cognitive processes of analogical alignment~\cite{gentner2017analogy}: ``a process of establishing a \textit{structured alignment}'' between two concepts that involves ``both alignment and projection.'' In projection, the base concept introduces some information into the target domain. In alignment, the process of attempting to \textit{align} two conceptual domains requires producing common structure(s) they share and may ``reveal commonalities that were previously not obvious in either'' domain \cite[p. 673-4]{gentner2017analogy}. 

As \Cref{fig:signwriting} shows, linking and grounding metaphors can drive the evolution of a notation. However, introducing new linking and grounding metaphors may require that a notation loosen or break its analogical alignments with prior metaphors. %
This loss of the ``coherence and inferential power''~\cite{gentner2017analogy} that came with prior alignments allows for leveraging new linking and grounding metaphors with new possibilities for coherence and inferential power.

Consider chemical formulas. 
Berzelius attempted to apply algebraic notation to chemistry~\cite[p. 10]{klein2003experiments}, inventing chemical formulas like $\mathrm{CuO}+\mathrm{SO_3}$ (originally expressed as {\small $\mathrm{CuO+S\stackrel{3}{O}}$
}) ---making a \textit{linking metaphor} to algebra (i.e., Berzelius thinking, \textit{Chemistry is Math}). %
In the process of analogical alignment, some properties of algebra's notational system mapped well onto the new situation---e.g., commutativity of $+$ and benefit of parentheses to group operations---while others, like $\times$ and associativity, broke semantically, whether through violation or loss of concrete meaning. The shorthand of capital letters for periodic elements was another linking metaphor and design choice: Dalton's alternative notation did not use letters, instead grounding the notation in a visual metaphor of atoms as indivisible balls visualized as circles of the same size~\cite[p. 35-40]{klein2003experiments}.

The introduction of new grounding metaphors can break aspects of linking metaphors (\ref{linking-metaphors}). Labanotation emerged from a linking metaphor to sheet music notation, but had a ``breakthrough'' when a co-inventor suggested rotating the horizontal staffs to be vertical: ``instead of a series of crosses written horizontally... %
vertically placed columns could index body parts, allowing symbols for movement to be written contiguously''~\cite[p. 278]{watts2015benesh}. Thus, %
the notation could more closely map to the left-to-right orientation of the body (e.g., with left arm gestures on the leftmost column). But the 5 staff-line channels were still kept as 5 columns---a residual artifact of sheet music notation.
A similar breakdown happened for SignWriting~(Fig.~\ref{fig:signwriting}). Sutton shows the music staff and bar line gradually receding: first the staff becomes three lines rather than five, then it is dropped entirely (``Who needed those unnecessary lines?'' \cite[p. 34]{sutton2015history}). Then, as the Deaf community began using the notation more and more, ``they were naturally stacking the symbols [vertically] to look like the human body,'' as it ``mimics the way signs look in real life'' \cite[p. 37]{sutton2015history}---applying a new \textit{grounding} metaphor that, by attending to spatial closeness of mapping, broke from artificial linearization. A final move to write in vertical columns fully eliminated the sole syntactic remnant from sheet music notation.

The analogical projection process %
can cause unique dimensions of the target domain---that have no correspondence in the base domain---to be dropped or temporarily forgotten, i.e., \textit{less cognitively salient as a non-alignable difference}~\cite{gentner97}. For instance, %
in dance, the positional arrangement of multiple dancers throughout a dance piece can be important, but Labanotation, by inheriting sheet music notation, focuses on performances of an \textit{individual} dancer divorced from space. %

\subsubsection{\cognitive{} \social{} \bf Incubation: Rapid iteration through application to diverse scenarios to achieve robustness}\label{pat:incubation}  %
Some early histories, such as for Feynman, Gruber (Markdown), Laban, and Sutton, resemble a process of incubation akin to iterative design. %
For instance, Laban and his colleagues worked tirelessly for about a decade before %
Laban ``demonstrate[d] his system to enthusiastic supporters'' to market, finance, and disseminate it~%
\cite[p. 279]{watts2015benesh}. Modern notations like programming APIs are incubated by inventors or small communities through rapid application and iteration to diverse usage scenarios. Not all notations have long incubation periods, however: math symbols, for instance, are the result of incremental suggestions proposed and selected over time~\cite{cajori1993history}. %

\subsubsection{\constraint{} \bf The agency of material constraints}\label{pat:material_constraints}

Notations are expressed on and through various media: pen, pencil, and paper; keyboard, mouse, and screen; whiteboard, stylus, and eraser. As discussed in prior work~\cite{blackwell2001cognitive, arawjo2020write}, these material constraints play deciding and constraining roles in initial notation development, circumscribing what kinds of representations are most easily expressed and which are resisted. For instance, the FORTRAN language was built around the limitations of the IBM business keyboards of the time~\cite{arawjo2020write}; adoption of the Flexowriter, which had sub- and super-script shift keys, might have meant array indices were accessed not as \texttt{a[i]} but as $\texttt{a}_\texttt{i}$. 

Latency and interaction also matter. People co-creating a shared language tend to converge on more abstract forms when they can simultaneously interact versus take turns: ``With even a very minimal level of graphical interaction (e.g., a tick to indicate comprehension), partners' drawings converged and developed from iconic to arbitrary signs... %
Without such interaction, drawings remained iconic''%
~\cite[p. 484-5]{galantucci2012experimental}. This is corroborated by evidence that comparison drives relational abstraction; even seeing two things in series without being prompted to compare will cause adults to miss shared abstractions~\cite{gentner2017analogy}.

\subsubsection{\constraint{} \bf Avoiding unintended symbolic collisions with pre-established notations}\label{pat:symbolic_collisions}

Other times, decisions are made to anticipate cultural ``collisions'' with the syntax and symbols of other pre-existing notational systems; you could call it ``cultural cross-mapping.'' If a new notation looks too similar to another well-known notation (at least well-known among intended users), it could cause confusion over interpretation: readers accustomed to the other notation could systematically misread the new one. For instance, Church hinted that his use of the $\lambda$ symbol arose from anticipated collisions with other meaning-saturated Greek letters; when asked why he chose $\lambda$ over other presumably less used Greek letters, %
he remarked: "eeny, meeny, miny, moe"~\cite{scott2018looking}.

\subsection{Dispersion and divergence (Stage 2)}

After the invention and incubation period, a notation may disperse to other communities and users, who re-interpret, amend, and take ownership of it. Historically, this typically follows an initial publication in the form of a book, academic journal, or mass media release. As users apply the notation to different situations in their unique contexts, the notation diverges from the original as ambiguities are settled through local debate or public contestation. Where multiple notations were invented for the same situation, %
``wars'' can emerge as competitors vie for attention and users. %

\subsubsection{\social{} \bf Inciting event where early adopters popularize and spread the notation}\label{pat:early_adopters} Like general innovations~\cite{rogers1962diffusion}, early adopters are critical to popularizing, spreading, and formalizing the notation. Consider the spread of Feynman diagrams, recounted by Kaiser \cite{kaiser2019drawing}. %
For Feynman, his initial notation was informal: notational rules, such as all vertices requiring exactly three edges, were not explicitly written down, and he only intended to use the notation himself as a calculating tool. It was actually Freeman Dyson who refined the theory and published it first~%
\cite[p. 175-195]{kaiser2019drawing}. %

\subsubsection{\social{} \bf As new communities take ownership of the notation and augment, interpret, and apply it to their unique context, the notation diverges.}\label{pat:divergence} Once a notation leaves its original inventor or community, ambiguities in interpretation and applications to new or unseen situations produce divergences from the original. These can be divergences in representation, but also interpretation. Circumstances where communities of notational practice become isolated (e.g., during wars) can cause divergences that may never be reconciled~\cite{watts2015benesh}. 

Undergirding divergences can be controversies between belief systems. For example, Kaiser recounts many geographically-bound divergences for Feynman diagrams and traces alternative beliefs as one cause~\cite{kaiser2019drawing}.

Divergence appears to be a natural process~\cite{galantucci2012experimental} and not strictly negative. %
Some notations, like Penrose diagrams~\cite{wright2013origins}, have a primarily educational character and can remain circumstantial and suggestive. Other times, however, divergences lead to controversy. %
Gruber's original post about Markdown ``does not specify the syntax unambiguously,'' leading to divergent interpretations when implementing a parser: %
``The best current way to resolve Markdown ambiguities... %
is Babelmark 3, which compares the output of 20+ implementations of Markdown against each other to see if a consensus emerges''~\cite{commonmark}. 
Another example is Labanotation---Laban used the term \textit{kinetography}, while \textit{Laban}otation was coined by Guest, who adapted the notation in the U.K. and U.S. Based on what school of dance notation one comes from, the exact same visual representation can be read in different ways%
~\cite[p. 284]{watts2015benesh}. 

\subsubsection{\social{} \formal{} {\bf Degree of formalization desired depends on need and context}}\label{pat:formalization_degree} %
How unambiguous a notation can be concerns the purpose of the notation and the degree to which formalization is beneficial to that purpose. Notations may be purposefully left ambiguous around dimensions of variation deemed less important or irrelevant. For instance, for music, dance and theatre, the notes merely serve as a script, but some channels, like volume and timbre, may not be explicitly encoded and up to performer interpretation%
~\cite{goodman1976languages}. %
For Markdown, Gruber felt that its success was ``\textit{due to}, not in spite of, its lack of standardization''~\cite{taskade-markdown-history}. %
By contrast, computer code %
cannot entertain any room for error, because the \textit{need} to build upon systems successively %
requires rigid determinism. Programming language communities thus create specifications %
to coordinate the ``correct'' performances (implementations). %

\subsubsection{\cognitive{} \bf Categorical extension: Extensions of notation \textit{within} an existing perceptual channel.}\label{pat:categorical_extension} 

Each notation leverages perceptual channels of human visual and embodied cognition (\ref{perceptual-channels}). Channels are a dimension of variation with concrete values it can take on---typically, categories mapped to meaning. Notations can evolve through extension of categories \textit{within} existing channels. For instance, Minkowski spacetime diagrams used straight lines to designate trajectories. An extension by Penrose introduced ``wavy'' lines within the same channel, as traveling ``observers and particles... breaking the convention that objects at the speed of light travel at 45 [degrees]''~\cite[p. 114]{wright2012advantages}. Suddenly, a line's \textit{wavyness} $\sim$ had meaning. %
Feynman diagrams have been extended in a similar categorical extension of trajectories, where gluons are expressed via spring-like squiggly lines.

Important here is the notion that a notation \textit{creates} categorical channels that can be later extended. For instance, Leibniz introduced $\int$ for both summation and integration~\cite[p. 36, vol. II]{cajori1993history}, which leveraged an unused channel of size (the symbol is taller than surrounding equation symbols)~\cite{euler1755institutiones}. Euler later changed the symbol to $\sum$, which Cauchy extended~\cite[p. 61, vol. II]{cajori1993history} to a vertical stacking channel (syntax for $n$ and $i$ placed above and below): $$\sum_{i=1}^n x_i$$ Making a categorical extension, Gauss leveraged the now-established channel for $\sum$ to introduce the product symbol $\prod$, in the process generalizing the concept of a looped operation to multiplication.

\subsubsection{\cognitive{} \bf Channel extension: Extensions of notation to \textit{unused} perceptual channels} \label{channel-extension}

As people take up notations, they extend or adapt the notation to suit their unique circumstances of use. However, the more socially popular and ``settled'' the notation, the more resistance one will anticipate facing when amending existing syntax and perceptual channels like ``shape'' and ``size.'' %
Modifications to existing perceptual channels %
risk unforeseen, emergent inconsistencies or conflicts (a similar phenomenon to the hesitance of software developers to make ``breaking changes''~\cite{brito2020you}). To cope with the risk of amending an existing channel, users of the notation perform \textit{channel extensions}---looking for previously unused or under-utilized perceptual channels and saturating them with meaning. %

For instance, on a whiteboard, one may sketch a UI in a single color, but thereafter realize they want to denote layout constraints and may use a second color to reduce confusion; the second color then conveys "layout constraints" while the first conveys "interface elements." The introduction of the second color introduces a new dimension of variation~\cite{marton2014necessary} and implicitly \textit{gives color meaning}. %

Consider also Berzelius' molecular notation. In his later work, Berzelius began slashing lines through simplified formulas, which ``denoted two `atoms'{}'' of the compound~\cite[p. 176-7]{klein2003experiments}---i.e., extending to an unused perceptual channel, a gestalt of continuity and overlap. %
He also used dots over elements to depict oxygen~\cite[p. 623]{berzelius1833betrachtungen}:

\begin{center}
    \includegraphics[width=0.4\linewidth]{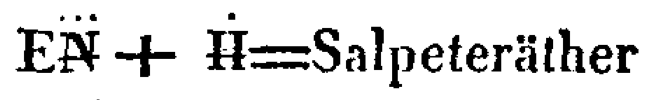}
\end{center}

Chemists later adapted Berzelius formulas to a graph notation to showcase structure: modern ``structural formulas.''
While molecular notation only tracked proportions of elements, structural formulae define how atoms are organized and connected, including features such as the quantity of lines designating the "strength" of the covalent bond (i.e., shared electrons). Lewis~\cite{lewis1916atom} introduced dot notation \textit{above} and \textit{beside} Berzelius' chemical symbols to denote bonds and valence electrons~\cite[p. 777]{lewis1916atom}: %

\begin{center}
    \includegraphics[width=0.95\linewidth]{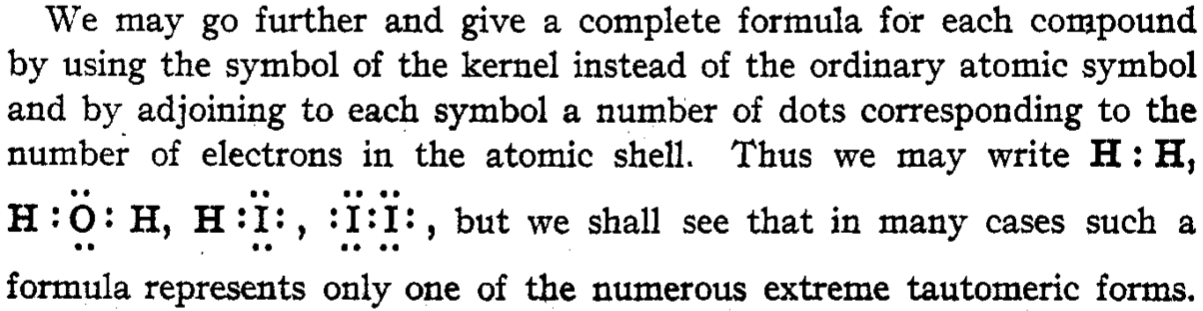}
\end{center}

(Notice here how Lewis adapted a previously made syntax, dots-above-letters, by tearing out its original semantics and replacing it.) Perceptual channels can be affordances~\cite{gaver1991technology}, leveraging  metaphors to embodied or cultural experience. In Coecke and Kissinger's adaptation of string diagram notation (\ref{grounding-metaphors}), filing off an end of the box allowed the viewer to keep track of  180 degree rotation where they could not before---saturating an unused perceptual channel of \textit{shape orientation} with meaning. In this way, a \textit{channel extension} attempts to avoid disturbing a notation's existing syntax and semantics.%

\subsubsection{\formal{} {\bf Informalizing moves: Augmentations informalize (unsettle) previously formalized (settled) syntax}}\label{pat:informalizing_moves}  

Ad-hoc extensions to notation can bring the notation back into a land of informality---an informalizing move---for the potential gain of theoretical power. %
An example can be found as Berzelius began extending his own chemistry notation, after it had become popular in the scientific community. In adding new elements such as the slash-through (\ref{channel-extension}), Berzelius made a note to separate out the previously-canonical (non-augmented) notation as an "empirical" formula—``they follow immediately from a correct analysis, and are immutable''—while his ad-hoc extension he calls a ``rational'' formula that is ``intended to give an idea'' of a concept and could ``vary according to perspective'' of the chemist but it was unclear which was the ``true'' formula~\cite[p. 177-8]{klein2003experiments}. Thus, while so-called ``empirical'' formulas had a purely \textit{descriptive} function, ``rational'' formulas sought to be \textit{generative}---to generate theories of reactions and chemical composition that may or may not hold up under later experiments. Berzelius was unsure how well his extensions aided the crafting of theories aligned with empirical evidence.

\begin{figure*}
    \centering
    \includegraphics[width=0.95\linewidth]{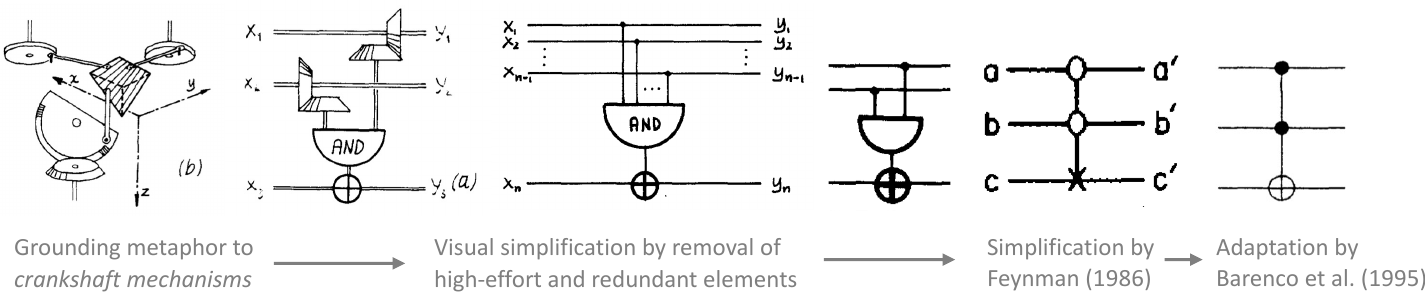}
    \caption{Origin of Toffoli gate notation in quantum circuits, revealing a grounding metaphor to crankshaft mechanisms by Toffoli~\cite{toffoli1981bicontinuous}, which was then adapted and simplified by Feynman~\cite{feynman1986quantum} which, in turn, was adapted by Bancelo et al.~\cite{barenco1995elementary}. Images  reproduced from original texts~\cite{toffoli1981bicontinuous, feynman1986quantum, barenco1995elementary} under fair use for purposes of academic analysis.} 
    \label{fig:quantum-circuit}
    \Description{"Origin of Toffoli Gate Notation in Quantum Circuits": A horizontal progression showing four stages of quantum circuit notation evolution, from left to right, with arrows connecting each stage. Stage 1: Shows the original Toffoli (1986) notation with a complex mechanical diagram resembling crankshaft mechanisms, featuring circular and linear elements with connecting lines. Stage 2: Shows visual simplification by removal of high-effort and redundant elements, streamlining the mechanical appearance while retaining core functional components. Stage 3: Shows further simplification by Feynman (1986), reducing the notation to essential geometric elements - circles, lines, and connection points. Stage 4: Shows the final adaptation by Barenco et al. (1995), presenting the modern quantum circuit notation with clean horizontal lines, control dots (filled circles), and target gates (circles with plus signs or X marks). The figure illustrates how a grounding metaphor to crankshaft mechanisms evolved through successive simplifications into the clean, abstract notation used in modern quantum computing.}
\end{figure*}

\subsubsection{\cognitive{} \bf Original semantics is reworked or discarded, while keeping a similar representation.}\label{pat:semantic_reworking}
Sometimes, the original \textit{semantics} may be entirely reworked while keeping a similar representation. For instance, Feynman diagrams have greatly diverged from their original semantics, but still remain useful: %
``With the failure of perturbation theory, one might have expected Feynman diagrams to lose their usefulness... %
Yet rather than toss away Feynman diagrams... %
many theorists began to exploit the diagrams in newfound, unprecedented ways... %
while discarding the original mathematical rules for their use''~\cite[p. 56]{kaiser2000stick}. SignWriting is another example, where Deaf users argued for the notation to be written from an egocentric (perspective of the signer), rather than an exocentric view (perspective of the observer). The syntax and lexigraphy remained the same, but the semantics diverged due to every sign now representing its mirror image~\cite{sutton2015history}.

\subsubsection{\cognitive{} {\bf Notation meets an un-representable circumstance and must be extended, amended, or a secondary notation develops.}}\label{pat:unrepresentable}
When the limitations of a notation's assumptions meet a situation that cannot be modeled (or whose salient features would be suppressed or cumbersome to read or write in the notation), it can cause the writer to either ``write around'' the notation (in the form of, say, writing in the margins or developing a secondary notation~\cite{blackwell2001cognitive}), amend or break it (at least one of its core tenets), or compel the search for a wholly alternative notation that better captures the obscured or un-representable information. For instance, one cannot encode emotion into logic notation, so authors try ``developing a representation language for the formalization'' of the new conceptual domain~\cite[p. 815]{lorini2011logic}.

\subsubsection{\social{} \formal{} \bf Gaining credibility and accelerating formalization by translation to already-formal proxy notation}\label{pat:credibility_translation} How a notation is granted a formal social status is often accelerated through mapping the notation to an already-recognized, well-established ``formal'' notation, which ``proves'' the new notation is similar or equivalent in power in some way. The formalization of the new notation can be ``bootstrapped'' from the existing one through the translation to the past notation, whether computationally (e.g., transpiling) or rhetorically (through proofs in academic work showing, for instance, ``equivalence'' between notations). A good example is the notion of Turing completeness: to more easily prove Turing completeness of a new programming language, one constructs a mapping from the new language to Turing machine notation or Church's $\lambda$-calculus. %
In programming languages, official ``spec'' documents produced by governance committees for language management, such as TS39's ECMAScript Specs, accomplish formality through translations to other notations such as formal grammars. A new notation that borrows from old ones might also accelerate the impression that it has a sufficiently formal character: Rust is a "formal" programming language that borrows from C, easing its credibility and adoption among C and C++ programmers. %

\subsubsection{\social{} \bf Notational wars between competing alternatives, mediated by usability and social power.}\label{pat:notation_wars}
Klein described the process of deciding upon a notation for molecular chemistry as a ``war'' \cite[p. 27]{klein2003experiments} where winning signs persist and losing signs die, in a kind of natural selection. In mathematics, Cajori presents many situations of competing symbols, such as between Leibniz and Newton's calculus notations~\cite{cajori1993history}. Notations that ``lose'' the war are relegated to history or stick around with less favorable status. Wars can also be resolved through deliberate efforts at unification and standardization (beginning of Stage 3). 

Whether a notation ``wins'' the war is not just a matter of usability, but of power. For instance, the equality symbol in mathematics $=$, introduced by Recorde in 1557, contested with Descartes' alternative symbol, resembling $\propto$ flipped horizontally. %
Mathematicians variously adopted either symbol, largely based on geographic proximity to Descartes. Cajori argues that the final choice of $=$ for equality was a matter of power, namely, Leibniz's prestige to adjudicate~\cite[p. 301-6, vol. I]{cajori1993history}. %

Similarly, Feynman and von Neumann were celebrated scientists whose notoriety could more easily sway adopters. %
Powerful educational institutions like Princeton were key mechanisms for ensuring the notation's spread, longevity, and vitality, winning out over Koba and Takeda's ``transition diagrams''~\cite{kaiser2019drawing}. %

\subsubsection{\cognitive{} \constraint{} \bf Routinization: Visual simplification and abbreviation of frequent arrangements}\label{pat:routinization} %

As they spread and are regularly used, notations can be simplified, whether visually (to forms that appear more abstract and require less effort to write), or through shorthands---new notations that compress information and effort of writing (the latter dependent on the medium). Examples include Einstein summation notation and how the $+$ symbol emerged from compression of effort to writing ``and'' again and again in early algebra~\cite{cajori1993history}. Figure~\ref{fig:quantum-circuit} shows how Toffoli gate notation in quantum circuits was simplified through successive adoption. %
Experimental semiotics suggests that abstract and simplified forms---in terms of visual dissimilarity of a symbol from its referent---emerge most strongly and quickly when people are co-located and converge as the number of users increases, as in a tight community~\cite[p. 486]{galantucci2012experimental}. This suggests that as a discipline organizes around a notation, it may increasingly abstract the notation to reduce writing effort and remove invariance, %
which at the same time could make it more difficult for newcomers to learn.

\subsubsection{\cognitive{} \social{} \bf Advances in conceptual understanding spurs notational evolution}\label{pat:conceptual_advances} A new semantic dimension of the source domain may be discovered (e.g., in science) or originally overlooked (whether due to cultural specificity or simplifying beliefs) during original notation's creation, spurring extensions or new notations. For example, the discovery of the electron prompted Lewis' extension of Berzelius' molecular notation to valence electrons and covalent bonds~\cite{lewis1916atom}. %
SignWriting deviated from Stokoe and Hamburg notations by acknowledging meaningful dimensions of variation in sign languages that had been marginalized: ``These systems are based on a set of symbols that represent the four `core' parameters... (or \textit{at least those considered appropriate by Stokoe and his colleagues}...) ... %
[B]ody postures and facial expressions, and the use of the gaze, although essential... %
[are] seldom taken into account" ~\cite[p. 118; emph. added]{bianchini2012writing}.

\subsection{Institutionalization and sanctification (stage 3)}

\subsubsection{\social{} \formal{} \bf Committees and institutions are established by notation developers and users to counter-act notation divergence by unifying competing alternatives and/or standardizing canonical interpretations and practices}\label{pat:institutions} As a notation spreads, it naturally diverges in usage and interpretation \textit{unless} social and material infrastructure is established to clarify and \textit{enforce} canonical representations, interpretations, and usage. We thus say a notation becomes ``formalized''---extended, refined, adjusted, and aesthetically stabilized. Adopters form communities which set out to define the proper and valid performances of a notation. The purpose of this formalization is to ensure \textit{reproducibility of action} across space and time: to produce \textit{unambiguous interpretations that lead to deterministic actions} which can then be \textit{judged} as valid or invalid \textit{performances} of the notation.\footnote{Here, we were inspired by \citet{goodman1976languages}'s usage of the term \textit{performance} in the context of notations.} %
As interpretation is a mediator of action, the ultimate goal is the coordination of action, whether human movements or mechanical ones.\footnote{For example, mathematical equations, understood through the backdrop of cultural rules, circumscribe a set of valid mathematical performances. The usage of the notation, the “manipulation” of symbols, is a performance under the constraints of the “rules.” Valid proofs are repeatable performances. %
Performing correctly is a practical and social achievement---for a notation can be misapplied by a novice.}

For Markdown, the divergences in interpretation of Gruber's original specification compelled power users to come together to create \textit{CommonMark}---```a standard, unambiguous syntax specification for Markdown, along with a suite of comprehensive tests'' to validate implementations~\cite{commonmark}---i.e., unifying notations into a standard and then enforcing it socially and materially. %
Another example are Python Enhancement Proposals (PEPs), ``the primary mechanisms for proposing major new features'' to Python''~%
\cite{pep1}. Language consortia resemble governments and law-making processes, referring to themselves as a ``formal governance process''~\cite{pep13}. %
Even data visualization notations have a history of committees~%
\cite{friel2001making}.

Efforts at notation unification and standardization through institutionalization can fail. ALGOL is a famous example from computing history. The final meetings fell apart in part because international collaborators, conducting work in local groups and  meeting intermittently with different people at each meeting, could not coordinate on one notation \cite{naur1968successes, history_algol}. %

\subsubsection{\formal{} \bf Formalization proceeds by community members suggesting, critiquing, and making persuasive arguments for amendments and extensions, which are then judged variously by users.}\label{pat:persuasive_arguments} %
Once a notation is in wide use, community members, including inventors themselves, cannot simply amend the notation, but must put forth arguments for revision if they wish to be ``formally'' accepted by the social group. Persuasive arguments can mobilize usability dimensions, such as effort and ease of writing, avoidance of cultural collisions, readability, closeness of mapping~\cite{blackwell2001cognitive}, and even appeals to community values (e.g., the Zen of Python~\cite{pythonic}). In computing, modifications often take the form of request for comments (RFC), persuasive documents around which the community debates a potential feature.%

\subsubsection{\constraint{} \bf Notation is materially institutionalized and reified through software and hardware}\label{pat:material_institutionalization} Once a notation becomes popular, it will be reified in materials such as machines and typesetters: sheet music notation are reified in software where its formal rules can be enforced and imposed on users~\cite{cavez2025euterpen}; UML notation becomes interactable in Visual Paradigm software, where it can generate code from the diagram~\cite{visualparadigm}; tables, used since antiquity~\cite{campbell2003history}, are reified in spreadsheeting software.

\subsubsection{\social{} \bf Notation is socially institutionalized, becoming a rite of passage and expected medium of knowledge transfer (e.g., exams, published papers)}\label{pat:social_institutionalization} %
The coordination of a ``proper'' or ``canonical'' interpretation is also socially accomplished through educational and other paratextual materials %
that teach and socially align how the notation should be \textit{correctly interpreted}, with canonical usage reinforced by power structures. For instance, a physicist's correct ``performance'' of Feynman diagrams in a job talk is tied to their ability to get the job. %

\subsubsection{\cognitive{} \social{} \bf Sanctification and subsequent devaluation of hard-or-impossible-to-capture dimensions of variation}\label{pat:sanctification} Highly popularized and formalized notations can become sanctified---what once sought to describe the world, becomes the schema through which users read the world, filtering what is `meaningful' or not. As a culture emerges around a notation, the notation may hold a near religious significance for its users. %
Reflecting on how music theory revolves around sheet music notation, Lilliestam remarks: ``Notation sets a framework, presenting a composer with both possibilities and limitations... [S]ound and timbre, micro intervals and `blue notes' and rhythmic subtleties cannot be captured by [sheet music] notation''%
~\cite[p. 198]{lilliestam1996playing}. Dimensions of variation in the world deemed irrelevant to Western classical music are hard or impossible to capture. He calls this \textit{notational centricity}---``the fact that we equate `music' with notated music... [and] the idea that notation contains the final truth about music''---which then causes musicologists to devalue un-representable dimensions~\cite[p. 196]{lilliestam1996playing}. %

Related to this is a need to continually clarify the ``correct perspective'' necessary for interpretation or usage, whether spatial, embodied, or a theoretical or philosophical belief, which may be leveraged in debates over the suitability of an extension to the notation. For instance, the Zen of Python is an official PEP, with lines such as ``There should be one--and preferably only one--obvious way to do it.'' These subjective principles, nonetheless outside the `formal' definition of the notation, form a common underlying culture that later Python usage, language extensions, and even users themselves are evaluated against~%
\cite[p. 9]{pythonic}. %

\subsubsection{\social{} \constraint{} \bf Changes are costly (financially, materially, socially) and propagate slowly.}\label{pat:costly_changes} Because notations have become socially and materially institutionalized, propagating changes faces considerable  resistance on both fronts. For instance, SignWriting was originally written from the perspective of the observer, not the signer. Yet in 1984, ``two Deaf staff members'' emphasized that ``we are expressing our own language from our own perspective. We see our own hands when we write, not someone else's.'' Sutton writes: ``They were correct. But it was an adjustment, because every textbook and every document published in SignWriting for the preceding ten years had been written receptively... The change took more than four years to complete'' \cite[p. 37]{sutton2015history}. An infamous example in computing is the transition from Python 2 to 3, where breaking changes caused such a strong ``resistance to change'' that ``Python developers began `back-porting' selected features'' to resist updating~\cite[p. 765-8]{malloy2019empirical}.

\subsubsection{\social{} \formal{} \constraint{} \bf Users fork the notation unofficially due to the high cost of direct amendment. Forks may eventually inspire or merge again into the ``official'' notation.}\label{pat:forking} To avoid the resistance of an official amendment to the notation, 
users may opt to circumvent the lengthy process altogether and instead create their own edition of the notation, unsanctioned by the governance regime. The prominence of transpiling in JavaScript ES6 circa-2010s is a good example: web developers would write code with the latest features to the language, before new amendments were propagated to updates in material infrastructure (browsers). These ``unofficial'' adaptations can then inspire developments in the ``official'' notation; for instance, CoffeeScript (a Python-syntax-like language that compiles to JavaScript) inspired Brendan Eich, the inventor of JavaScript \cite{eich2011harmony}. The TC39 consortium responsible for formally extending the language then adopted extensions including arrow functions, optional chaining (\texttt{?.}), \texttt{class} and \texttt{super}  \cite{eich2011jsconf}. %

\subsubsection{\social{} \formal{} \bf What begins as secondary notations or conventions around the base notation can be formalized, either becoming notations themselves with rigid rules of practice, or adopted into the base notation.}\label{pat:secondary_notations} Python's formal inclusion of whitespace into the language itself (an idea adopted from ABC) is an example of bringing informal syntactic practices utilizing an undefined perceptual channel into the formal semantics of the language (assigning meaning to whitespace). Similarly, what starts as secondary notations or mere ``conventions'' like docstrings~\cite{pep257}---which a user may choose not to follow without penalty---can later become formalized when exactness is required by downstream parsers, such as tooltips in software IDEs which rely upon an expected grammar to typographically render the docstring. %

\subsubsection{\constraint{} \bf Re-encodings: Notation suited to one medium are re-encoded to fit the affordances and constraints of another}\label{pat:re_encodings} %
Once a notation proves useful and becomes institutionalized, this is not the end of its development. It often must fit into new material constraints, especially prompted by the limits of computer systems and their standardized encoding schemes and input devices. SignWriting symbols must be added to the Unicode Standard; %
Feynman diagrams must be described in FeynArts; electronic circuits are encoded in SPICE netlists; chemical notation is linearized in SMILES (Simplified Molecular Input Line Entry System) notation for electronic search and storage (right is SMILES~\cite{ucalgary_smiles_tool}):

\begin{center}
    \includegraphics[width=0.6\linewidth]{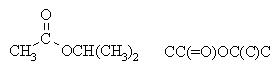}
\end{center}

Although these are linearized encodings constrained by cultural standards around ASCII and the Latin keyboard, technological developments can relax previous material constraints; for instance, %
how advances in pen-input devices have inspired mixing typewritten programming with handwritten diagrams~\cite{codeshaping, arawjo2022notational}.

\subsubsection{\social{} \bf Inventors lose control: initial intent is bent and broken as notations are taken up and formalized by wider communities, who grow to have different values and commitments}\label{pat:lose_control} A surprising pattern shared across histories is where the original inventor(s) seem to ``lose control'' of the notation. Once a community takes ownership, its values and intentions diverge from the original inventors, sparking controversies that can end with the inventors leaving communities, disowning new incarnations or reasserting control. Examples include how Guess's adoption of Laban's kinetography diverged from Laban's, causing Laban to write a lengthy treatise to spread the ``real'' version~\cite{watts2015benesh}; how Gruber became ``very upset'' when people assembled to standardize Markdown, which led to him trying to police how the term is used~%
~\cite{atwood2014commonmark}; and Guido van Rossum stepped down as Python's ``benevolent dictator for life'' after a heated clash over PEP 572: Assignment Expressions~\cite{pep572}: %
``I don't ever want to have to fight so hard for a PEP and find that so many people despise my decisions''%
~\cite{vanrossum2018transfer_of_power}.

\section{An Emerging Theory of Notation Development} \label{sec:unified-theory}

Our historical analysis suggests that, cognitively and socially, a notation proceeds by:
\begin{enumerate}
    \item Enumerating dimensions of \textit{meaningful variation} in the target domain, which proliferate as more situations are encountered or considered (whether by inventors or users)
    \item Mapping dimensions of meaningful variation to perceptual channels of representation
    \item Designing the notation to leverage perceptual affordances by visual analogy to embodied transformations like pouring cups or rotating shapes, and ensuring these ``natural'' manipulations hold meaning in the target domain
\end{enumerate}

Notation definition therefore proceeds whereby \textbf{a dimension of meaningful variation in the \textit{conceptual/empirical} domain is mapped to a dimension of \textit{perceptually distinguishable} variation in the \textit{representational} domain}. What counts as \textit{meaningful variation} is determined through experiment, cultural knowledge, or simply free choice. A dimension of variation previously unknown or thought to be invariant may become meaningful when further empirical evidence suggests it (e.g., the discovery of electron bonds prompting Lewis' extension to molecular notation~\cite{lewis1916atom}). A notation abstracts away from iconic representations by stripping away dimensions of invariance, as invariant dimensions take effort to write but provide no additional information.

Alongside the \textit{social} stages of notation development above are three \textit{functional} stages that emerge from reflection upon our analysis---\textit{descriptive}, \textit{generative}, and \textit{evaluative} stages (borrowing terminology from Generative Theories of Interaction~\cite{beaudouin2021generative}):

\begin{enumerate}
    \item \textit{Descriptive} stage: People seek to describe world or conceptual phenomena to manage complexity and/or coordinate action. %
    Notation development in this stage proceeds by the closeness of mapping~\cite[p. 330]{blackwell2001cognitive}, i.e., attending to how representable referents are in the notation. Development proceeds through application to diverse scenarios, where the notation is tested and amended for its (in)capacity to represent dimensions of meaningful variation in the target domain and users' corresponding capacity to apply and interpret it consistently.
    \item \textit{Generative} stage: Once a notation can sufficiently describe phenomena in a domain, humans use it as a generative theory: to generate new ideas, concepts, and hypotheses that they can test or implement. For instance, musicians compose music in notation without touching an instrument~\cite{cavez2025euterpen}; chemists conjecture about reactions based on deductive symbol manipulations~\cite{klein2003experiments}; HCI researchers look into the ``generation of new micro-gestures'' by ``build[ing] a tree of all possible glyph combinations'' in the notation, and identify gaps through comparison to a notated dataset of micro-gestures covered by existing literature~\cite[p. 16]{micro-gesture-notation-2023}.
    \item \textit{Evaluative} stage: If a notation achieves great popularity in a domain, it can become a metric through which empirical or conceptual phenomena are judged---the notation becomes a tinted lens through which the world is read. What began as a mapping from dimensions of variation of phenomena, to dimensions of perceptual variation, now is inverted, with  the ``represent-ability'' of phenomena in the notation filtering out what dimensions of empirical variation matter.\footnote{A plausible cognitive explanation for this stage is that notations form new schemata~\cite{tversky2000some}, which then ``shape what we see and hear, how we store that information, and how we access it for later use'' \cite[p. 921]{von1993inhibitory}. Through a series of experiments, Von Hippel et al. show that learned schemata ``inhibit the amount of information that is remembered by suppressing the perceptual encoding of both relevant and irrelevant information''—quite literally, %
    biasing users towards foregrounding dimensions that hold meaning in the schema, and overlooking dimensions that do not.} Not all notations may be sanctified in this way. For instance, some Deaf people are concerned that SignWriting could become an evaluative metric determining how new forms of American Sign Language are judged and valued~\cite{DeafPerspectivesOnSignWriting}.
\end{enumerate}

\paragraph{\bf A worked example}

To concretize early cognitive patterns and reflect on our own notational practices as HCI and software designers, we examine a notation that emerged during a collaboration on a different research project, in the context of human-AI grounding.

\begin{figure*}
    \centering
    \includegraphics[width=0.85\linewidth]{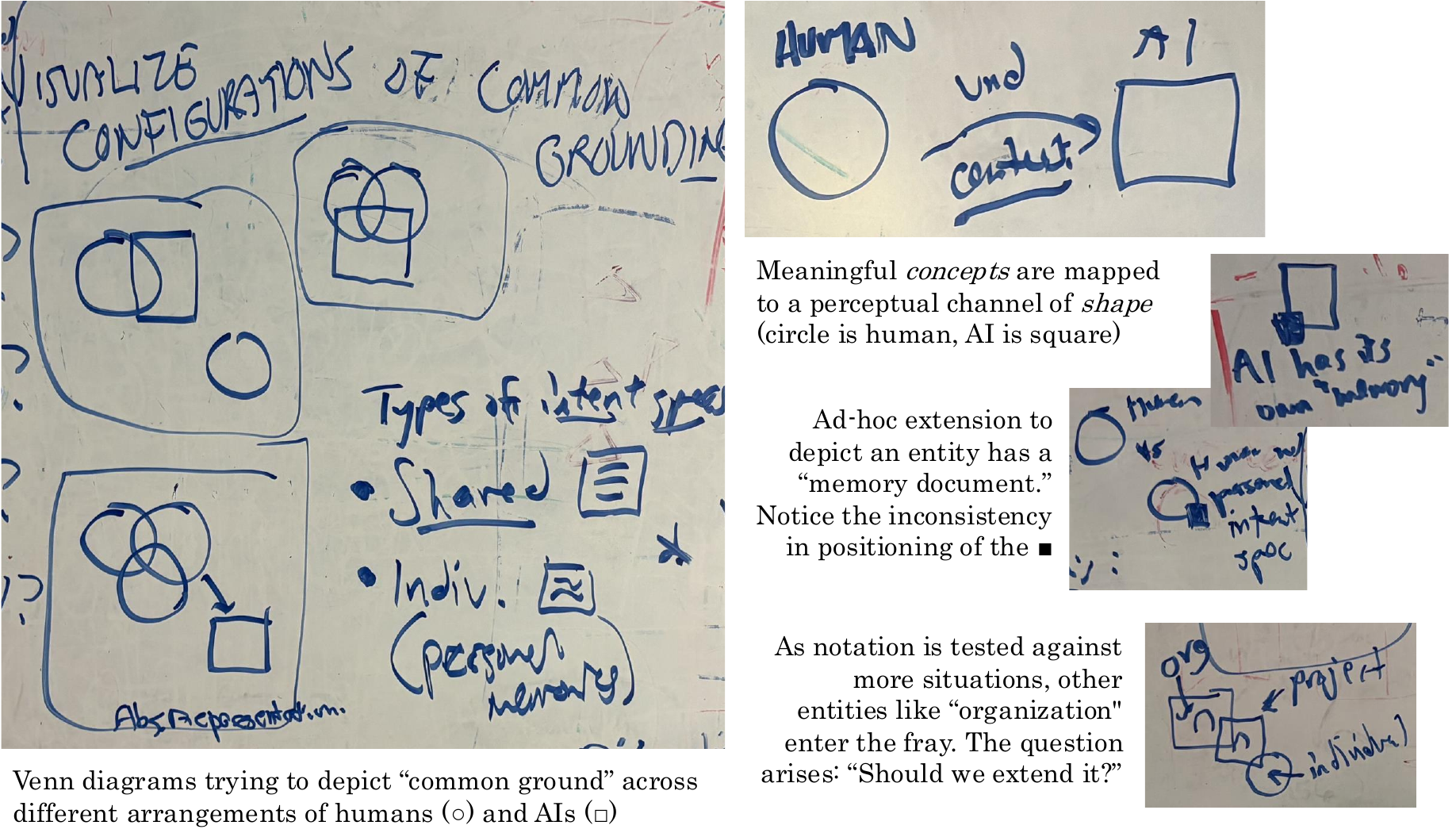}
    \caption{A notation in incubation: Excerpts from an HCI researcher's whiteboard. ``Common ground'' between humans and AI is visualized through a linking metaphor to Venn diagrams. During the analogical alignment process, dimensions of variation in the target domain are discovered to be un-representable or obscured, prompting notation extension, change, or abandonment.}
    \label{fig:common-ground-notation}
    \Description{Four hand-drawn whiteboard sketches showing the iterative development of a notation system for representing "common ground" between humans and AI agents. Top left sketch: Shows various arrangements of circles (representing humans) and squares (representing AI) in Venn diagram-like overlapping configurations, with handwritten labels indicating different entity types and relationships. Top right sketch: Shows a simpler version with one circle labeled "HUMAN" and one square, with an overlapping area and additional notational elements being tested. Bottom left sketch: Shows an attempt to add memory documents represented by small squares (■) positioned inconsistently around the main shapes, demonstrating the challenge of extending the notation to new concepts. Bottom right sketch: Shows further iterations exploring how to represent different organizational entities. The figure captures the real-time evolution of a research notation, showing how conceptual challenges emerge during analogical alignment between the source domain (Venn diagrams) and target domain (human-AI collaboration), leading to various ad-hoc extensions and ultimately questioning the adequacy of the original metaphor.}
\end{figure*}

Figure~\ref{fig:common-ground-notation} shows the common ground notation as a work-in-progress, created from a linking metaphor to Venn diagrams. The metaphor was initially made due to a structural similarity: the ``human'' and ``AI'' have separate ``understandings'' unknown to each other, but also share a common understanding (``common ground''~\cite{clark1996using}). Quickly we realized that we need to denote the \textit{type} of entity; thus, we extended Venn notation by extending a previously invariant perceptual channel of \textit{shape}---now, circle means ``squishy'' human, square means ``sharp'' AI (a \textit{grounding} metaphor). In doing so, we opened a perceptual channel of shape with two categories of representation that later collaborators might extend if the conceptual circumstances call for distinguishing a new entity (e.g., triangle for an organization). Beyond two overlapping shapes, Venn diagrams have the nice property of generalizing to broader arrangements of human and AI where multiple humans/AI are present and in different linkages (Fig.~\ref{fig:common-ground-notation}, left). 

However, in the process of use, we discovered that the analogy can mislead: the overlap creates an illusion that humans and AI share beliefs, when in reality, there is always a gap. It is also difficult to denote ``A's model of B'' and vice-versa, recursively (``A's model of what B thinks of AI''). AI agents also have properties that differ from humans, such as potentially never forgetting what was said. And, the notion of memory documents that could be shared between collaborators was lost in the shuffle. Is the overlapping area a memory document? What about information that the AI is collecting on human A's collaborator B, which is hidden to B? We considered adding a $\blacksquare$ to resemble a `document' in a bottom corner of an entity to denote that it has an external memory document, but this too leads to ambiguities and questions. Don't all and only AI agents need such memory documents (i.e., `isn't this an invariant?')? If so, why depict the documents explicitly at all? 

To summarize, the cognitive process of analogical alignment between Venn diagrams~\cite{venn1880diagrammatic} %
and human-AI grounding revealed semantic breakages that `didn't quite fit.' Dimensions of meaningful variation, here concepts from Clark's theory of common ground~\cite{clark1996using}, were not easily representable, causing us to either extend the notation, or abandon it entirely. But, the process was also productive. Adoption of Venn diagram notation formed a framing device, a schema or ``tool for thought'' that we then used to interrogate and clarify the dimensions of meaning that mattered.

\section{Preliminary Implications for Design} \label{sec:implications}

Here we present suggestions for system designers, with concrete examples inspired by our patterns. These are just some interesting ideas that came to mind, rather than an exhaustive list. %
While implementation-agnostic, many ideas could benefit from generative AI and LLMs' world knowledge to anticipate ambiguities, surface cultural collisions with other notations, or help users hone persuasive arguments for notation amendment.

\newcommand{\implication}[1]{\subsubsection*{\textbf{#1.}}}

\implication{Help identify, map, and prioritize the dimensions of variation in the problem domain} A system could help users collect examples from the conceptual/empirical domain, surface dimensions of variation, estimate their frequencies, and flag overlooked dimensions. This mirrors need-finding in user-centered design, where practitioners enumerate dimensions of user variation and needs.

\begin{whyquote}
\textbf{Why?} Notations map dimensions of meaningful variation in conceptual/empirical domains to perceptual channels (\ref{pat:dimensions_of_variation}). Once established, later introducing overlooked dimensions is challenging (\ref{pat:conceptual_advances}). Conversely, foregrounding invariant or rare dimensions wastes effort and can mislead users (\ref{pat:dimensions_of_variation}).
\end{whyquote}

\begin{examplequote}
\textbf{Example:} For instance, for $\mu$-glyphs~\cite{micro-gesture-notation-2023}, a system might have supported designers to review micro-gesture literature and help taxonomize the variation and dimensions of micro-gestures.
\end{examplequote}

\implication{Clarify which perceptual channels are still available, and which ones are saturated with meaning} While a notation is being created, users should be guided in the use of perceptual channels (e.g., colour, size, proximity, closure). By showing how these channels are used (``saturated'') by the notation, users can make more informed decisions: overused perceptual channels might benefit from being simplified, whereas unused channels might reveal opportunities. %

\begin{whyquote}
\textbf{Why?} Notations evolve by leveraging unused perceptual channels (\ref{channel-extension}) and by adding meaning to existing channels (\ref{pat:categorical_extension}). Some channels might be overlooked (\ref{perceptual-channels}).
\end{whyquote}

\begin{examplequote}
\textbf{Example:} In Figure~3, perceptual channels like ``shape'' and ``overlap'' are saturated, while others like color and size presently meaningless. A system might list these channels and their semantic mappings, and suggest introducing unused channels as new scenarios arise and new dimensions of variation appears. %
\end{examplequote}

\implication{Anticipate ambiguities, inconsistencies, and potential misinterpretations} Users should have some signals about the quality of the changes they are introducing to a notation. A system should flag inconsistencies or ambiguities and help the user resolve them. %
The system could identify elements that may be misinterpreted by others (e.g., the use of colors incompatible with red-green blindness, or the overlap of a symbol with another well-known notation).

\begin{whyquote}
\textbf{Why?} Similar-looking notations create confusion (\ref{pat:symbolic_collisions}), unaddressed ambiguities can lead to divergent interpretations (\ref{pat:divergence}), and incidental visual resemblance can mislead users (\ref{grounding-metaphors}).
\end{whyquote}

\begin{examplequote}
\textbf{Example:} Consider a user's hand-drawn game in a grid where players are allowed to place and ``move'' their ``units,'' X's and O's, to have them ``battle'' adjacent units. %
A system could warn the designer that the visual resemblance to Tic-tac-toe might cause confusion, and suggest alternative designs. The system might also prompt the user to clarify whether symbol size matters.  %
\end{examplequote}

\implication{Encourage simplifying designs that reduce the effort required to read and write the notation} %
Users should be able to discover simplifications and measure the complexity of their notations in terms of writing effort, memorization effort, and space efficiency. Systems should suggest simplifications, abbreviations for frequent arrangements, and facilitate the comparison of alternatives based on quantitative metrics (e.g., number of strokes needed).

\begin{whyquote}
\textbf{Why?} The goal of a notation is to manage complexity, reduce writing effort, and facilitate comparison (\ref{sec:manage-complexity}). Frequent arrangements are often abbreviated as notations evolve (\ref{pat:routinization}).
\end{whyquote}

\begin{examplequote}
\textbf{Example:} A system could store and compare example usages of a notation to identify critical dimensions of variation and frequent arrangements. This could suggest meaningless details to remove and shorthands for repeated patterns. The system could also predict writing and memorization effort.
\end{examplequote}

\implication{Help users make linking metaphors to existing notations, and support the resulting process of analogical alignment and mismatch identification} A system should help users search, discover, and make linking metaphors to existing notations. This analogical process should consider the source and target notations’ application domain, the perceptual channels leveraged, and the afforded operations. Crucially, structural misalignments should be highlighted with suggestions on possible solutions. 

\begin{whyquote}
\textbf{Why?} Linking metaphors to past notations is crucial during notation invention (\ref{linking-metaphors}), but the process of analogical alignment from source to target notation can cause users to overlook important dimensions (\ref{pat:break_alignment}) or invite misinterpretations (\ref{grounding-metaphors}).
\end{whyquote}

\begin{examplequote}
\textbf{Example:} %
In our work (Figure 3), Venn diagrams suggest overlapping regions represent shared content, which is misleading because the target domain involves perspectives that can never fully align. A system might have warned us that this structural misalignment could invite misleading conclusions. %
\end{examplequote}

\implication{Encourage the use of metaphors grounded in embodied experience} In addition to supporting linking metaphors to existing notations, grounding the notation in embodied human experience should be encouraged. The system should suggest ways to convey how to use the notation by using physical metaphors that invite certain actions. In doing so, the system should ensure suggested metaphors do not cause cross-mappings and other mismatches that invite forbidden actions or consistent misinterpretations.

\begin{whyquote}
\textbf{Why?} Grounding notations in embodied perceptual experience might lead to increased usability by leveraging user intuition (\ref{grounding-metaphors}). This needs to be done with care, as these metaphors can also lead users astray when the afforded actions do not correspond to meaningful operations in the target domain (\ref{grounding-metaphors}).
\end{whyquote}

\begin{examplequote}
\textbf{Example:} For a notation representing body movements, suggestions could visually or structurally resemble body parts but be simplified (abstracted) as much as possible to reduce effort. %
\end{examplequote}

\implication{Tailor suggestions to material constraints} The system should be aware of the type of media (e.g., pen and pencil with physical paper) that will be used to represent the notation and give constraints and suggestions tailored to that media. Additionally, the system could provide support for users wishing to encode a given notation in a new medium by suggesting adaptations that remain as consistent as possible to the original.

\begin{whyquote}
\textbf{Why?} Material constraints play deciding and constraining roles in notation development (\ref{pat:material_constraints}). The same notation may need re-encoding when moving between media (\ref{pat:re_encodings}).
\end{whyquote}

\begin{examplequote}
\textbf{Example:} %
For keyboard-typed notations, systems should suggest ASCII-compatible characters and estimate writing effort from character accessibility on a given keyboard layout.
\end{examplequote}

\implication{Support the flexible transition between using the notation and iterating on its construction} Much like programmers can work on the code and then run it to test it, users should be able to work on the notation and also regularly test it. Testing should happen throughout the development of a notation, and multiple users should be able to do so simultaneously. Systems should suggest ways to correct the issues identified during testing.

\begin{whyquote}
\textbf{Why?} Notation development historically resembles iterative design (\ref{pat:incubation}). These testing phases often lead to amendments to the notation (\ref{pat:unrepresentable}). Testing also facilitates collaborative efforts to evolve a notation (\ref{pat:material_constraints}).
\end{whyquote}

\begin{examplequote}
\textbf{Example:} Systems could offer a testing playground that enforces valid notation uses. %
The system could also ask clarifying questions when ambiguity arises or prompt edge cases: ``Can your notation represent [unusual scenario]?’’ and track scenarios that cause breakdowns or require extensions.
\end{examplequote}

\implication{Support personal variations of a well-established notation which can inspire the original notation} A system should support creating personal variants of notations that can disrupt or break prior formalization (akin to making software ``malleable''~\cite{litt2023malleable}). %
Similarly, authors of the canonical notation should be able to track and draw inspiration from variants.

\begin{whyquote}
\textbf{Why?} Users create unofficial variants of notations due to the high cost of direct amendment (\ref{pat:forking}). In turn, these variants may inspire developments in the original notation. Additionally, as notations are shared and adapted by others, they often diverge (\ref{pat:divergence}). Ad-hoc extensions can temporarily informalize aspects of an existing notation for potential theoretical gain (\ref{pat:informalizing_moves}).
\end{whyquote}

\begin{examplequote}
\textbf{Example:} Consider a user of Lua who misses higher-order features from CoffeeScript.\footnotemark{} A system could allow the user to introduce new syntax and semantics in-line---not as a macro but in an ad-hoc manner. The system thereby ``lifts'' Lua from the land of rigid formality---an ``informalizing move'' (\ref{pat:informalizing_moves})---making it more malleable in the user's local area and simulating its execution. %
\end{examplequote}

\implication{Facilitate the sharing, discovery, and critique of new notations} Sharing a notation should be easy so that others can use or modify it. Shared notations should be searchable to allow others to discover, adopt, critique, and ``fork'' them.\footnotetext{This situation resembles MoonScript, a scripting language that compiles to Lua and which was developed from the author's wish to write more concise code (avoid ``keyword noise'') with CoffeeScript-like abstractions: \url{https://moonscript.org/}.}

\begin{whyquote}
\textbf{Why?} Popularizing a notation is challenging. Early adopters are important to popularizing, spreading, and formalizing notations (\ref{pat:early_adopters}). Testing against diverse scenarios is critical to strengthening a notation's design, but designers are often limited by personal and cultural biases~(\ref{pat:dimensions_of_variation}, \ref{pat:incubation}).
\end{whyquote}

\begin{examplequote}
\textbf{Example:} Notations could use a standard file format to be easily shared on platforms that also support advanced search, using domain, material constraints, perceptual channels (e.g., ``use proximity’’), or grounding metaphors.
\end{examplequote}

\implication{Facilitate coordination of notation governance through discussion, consensus-building, and surveying mechanisms} The community around a notation might be supported in debating changes. %
Similarly, processes should exist for handling conflicts, disagreements, and passing leadership positions to future members. %

\begin{whyquote}
\textbf{Why?} Notation institutionalization involves establishing committees and governance structures to ensure reproducibility across space and time (\ref{pat:institutions}). Diverse usage contexts may drive a notation's evolution (\ref{pat:formalization_degree}). Communities may assume control over the notation's original inventor, causing conflicts (\ref{pat:lose_control}).
\end{whyquote}

\begin{examplequote}
\textbf{Example:} Platforms could support communities in discussing by defining roles (maintainer, contributor, user), surveying mechanisms to reach community consensus on proposals, or monitor usage to identify frequent patterns or under-utilized features. %
\end{examplequote}

\implication{Help users put forth persuasive arguments for requesting the revision of an established notation managed by a community} Users should be supported in articulating why the modification is needed by considering factors such as usability, effort, consistency, and adherence to community values and principles. Additionally, platforms should estimate the cost of a modification and suggest how to mitigate those.

\begin{whyquote}
\textbf{Why?} Once a notation is in wide use, community members cannot simply amend it but must put forth arguments for revision to be formally accepted (\ref{pat:persuasive_arguments}). Changes face resistance and are costly without mitigation plans (\ref{pat:costly_changes}).
\end{whyquote}

\begin{examplequote}
\textbf{Example:} For instance, a system might support users in preparing a Python Enhancement Proposal (PEP) to the Python language~\cite{pep1}. The system might retrieve similar debates and prior PEPs to guard against duplicate ideas. The system might also help the author tune designs and make appeals to community values, such as what is ``pythonic'' ~\cite{pythonic}.
\end{examplequote}

\section{Towards Dynamic, User-Defined, Incrementally Formalized Abstractions} \label{sec:discussion}

In this paper, we laid some groundwork towards a theory of notation development. %
However, charting the evolution of notation is a major undertaking that cannot be completed comprehensively in one paper. %
We hope our findings offer useful starting points for more robust theories of notation and abstraction development. In particular, our work is primarily situated at a macro-scale; we sketch out high-level patterns but omit detailed depictions of how formality is practically accomplished moment-by-moment. Future studies might enrich our social constructionist view of formality by tracing moment-by-moment (in)formalizing moves, perhaps connecting to research in experimental semiotics~\cite{galantucci2012experimental}.%

Our work contributes to a longstanding dream of dynamic abstractions in HCI, where users can dynamically communicate and express themselves through notations (interfaces) that they are most comfortable with at the moment of expression, beyond ones predefined by developers~\cite{vaithilingam2024imagining, victor2022, victor2024, masson2024directgpt, vaithilingam2024dynavis}. Such ``dynamic grounding'' \cite{vaithilingam2024imagining} promises to reverse the traditional unidirectional power dynamic of creativity support tools  \cite{li2023power}. Bret Victor stressed that dynamic grounding remains a key bottleneck to the future of HCI, reflecting that his advances with DynamicLand, though promising, have been~\cite{victor2022}: %

\begin{quote}
    ``%
    ...mostly within toolsets that had already been authored. We don't have anything like general-purpose \emph{de novo} computational sketching. Text-and-keyboard programming is too slow, demanding, and inward-facing to support a conversation about something other than itself... If we're able  to express a new idea, in realtime conversation, with a spontaneous hand-drawn dynamic model, that will be quite a milestone.''
\end{quote}

To conclude, we connect to the history of user interfaces and raise broader implications for systems supporting language evolution.

\subsection{User Interfaces as Notations}
Given that user interfaces are a form of notation, it is not surprising that their evolution closely follows the patterns we identified. 
The HCI community has recognized the importance of grounding metaphors and developed theories and principles to help interface designers find and develop such metaphors~\cite{nealeChapter20Role1997, hutchinsMetaphorsInterfaceDesign1987, carrollMetaphorCognitiveRepresentation1982}. 
For example, Reality-based interactions~\cite{jacobRealitybasedInteractionFramework2008} argues for interfaces grounded in embodied experience, such %
as our understanding of physics and body awareness.
Early interfaces were designed based on grounding metaphors, such as the desktop metaphor simulating an office, with files stored in folders and information displayed in windows~\cite{moggridgeDesigningInteractions2007}. Similarly, recent works like Textoshop make use of linking metaphors such that one interface, like Photoshop for images, is adapted to a different context or domain, like writing~\cite{massonTextoshopInteractionsInspired2025}.

However, \citet{blackwell2006reification} points out that metaphors have not always been a `slam dunk' in terms of promoting usability or utility in HCI design. Early interfaces evolved through use, constraints (e.g., the trashcan appearing on the desktop~\cite{moggridgeDesigningInteractions2007}) and by leveraging new perceptual channels (e.g., icons replacing text~\cite{smithPygmalionCreativeProgramming1975}) to the point that original metaphors are hardly recognizable anymore: few people know that the caret in a word processor is a metaphor for a typewriters' head or  that the way we select text is a metaphor of how we use a pen to stroke text~\cite{moggridgeDesigningInteractions2007}. In fact, some metaphors do not make much sense anymore for many new users. For example, many click an icon of a floppy disk to save their document, despite not knowing what it represents. 

Our work on notation evolution shows a similar pattern: as a   notation develops, it often breaks fundamentally from the metaphor(s) that inspired its initial creation. As time passes and users and usage changes, it often ceases to be a metaphorical reference to some other object. Freed from its metaphorical baggage, it
can be more freely modified (like "spaces" in the MacOS that have nothing to do with the properties of physical desktops~\cite{blackwell2006reification}). In other words, metaphors, or at least recognizable correspondences with pre-existing knowledge, may be valuable precisely \textit{because} initial correspondences are ultimately forgettable. When and where the initial family resemblance can be hard to make out, the new abstraction can be re-deployed as a new linking metaphor for future notations down the chain~\cite{lakoff2000mathematics}. 
In this way, we %
view `forgettability' as an inevitable consequence of notation evolution---even a sign of its success.

\subsection{How can future systems support notation \textit{evolution}, not just translation?}

Recent AI systems are incredibly effective at \textit{translation to existing formalisms}, but not so much the \textit{creation of new formalisms}. They move us `horizontally' along the square of formality (Fig.~\ref{fig:square-of-formality}), but say little, and can even impede, the creation and proliferation of new notations. As evidence of this, emerging research shows that LLMs stifle cultural evolution by reproducing and amplifying dominant languages~\cite{farina2025english, twist2025llms}. Software developers are now choosing programming notations not because of how well they like the notation, but how well it is  understood by LLMs.\footnote{E.g., ``The main reason I now work primarily in Python is that LLMs are better at it and default to it, rather than to R or Julia''~\cite{Rivest2025Tweet}.} Without a new paradigm that embraces language evolution, ``we risk being stuck forever in the software ecosystem of ~2020–2023''~\cite{Rivest2025Tweet}---and potentially other linguistic ecosystems as well.

In such a context, how might future information systems continue to support the natural evolution of language? %
Some emerging work on in-context learning considers how LLMs can produce flexible domain-specific languages, e.g., Grammar Prompting~\cite{wang2023grammar}. However, this work typically takes the new abstraction to be learned for granted, again hiding its genesis from view. %
While current AI systems support ``horizontal'' translations from informal ideas to established notations, how should we ensure that the ``vertical'' process of creation---new notations, new abstractions---is also supported? %
How might we co-create a new notation with a machine, and thereafter communicate through that notation, even share out the notation to broader communities? %

Our patterns and stages have some implications for these questions. First, it seems that new notations are primarily---although not only---created through linking metaphors to past notations. Notation evolution might be analogized to natural language evolution, then, where pidgin dialects become creole~\cite{bickerton2015roots}. Although up close, notations seem separate, %
at a distance, notation development appears more like an evolutionary process of gradual mutation with occasional breakthroughs, %
aligning with a social constructionist perspective on technology that views ``the developmental process of a technological artefact... as an alternation of variation and selection''~\cite[p. 411]{pinch1984social}. %

Second, our work reminds us that formalization is a social process. Formality is not a spectrum nor a reachable destination, but requires careful negotiation and always remains subject to contestation. For future systems that aim to support incremental formalization, treating formalization as a process full of debate and value-setting---rather than a strictly translational, technical, or rote task---will be critical. While we want our tools to evolve with and for us, we do not want sweeping notational changes to be forced-pushed without community consent.

Third, while many of the dimensions of meaningful conceptual/empirical variation remain \textit{implicit} to inventors, there is an opportunity for systems to make these assumptions more explicit and tangible. Beaudouin-Lafon's \textit{reification} is helpful here: ``a process for turning concepts into objects'' whereby ``[i]n user interfaces, the resulting objects can be represented explicitly on the screen and operated upon''~\cite[p. 449]{beaudouin2000instrumental}. How might future interfaces to support notation development reify dimensions of variation, rendering them tangible, even manipulable? %
What would ``swapping'' saturated perceptual channels---color for shape, say---mean? %

Fourth, although standardization committees and practices like RFCs are at least as old as computing itself, we cannot help but notice almost no HCI work focuses on supporting, say, the process of submitting a proposal to extend a notation such as a programming language. %
Work that also directly designs for notation \textit{unification} %
is noticeably absent. Fruitful here are empirical studies on language governance in software engineering~\cite{barcellini2009participation,keertipati2016exploring}, which offer more insights for targeted design. %

Finally, notation %
is also a matter for explainability and the continued flourishing of human cognition~\cite{tankelevitch2025tools}. For instance, if AI systems can begin to accomplish tasks that surpass human cognition that humans must try to understand and verify, the notations of the past may not be the best forms for communicating with humans. Both humans and AI might also benefit from new notations, even in existing domains; for instance, some research  shows new notations for probability theory and electricity improve learning outcomes for novices~\cite{cheng2002electrifying, cheng2011probably}. We hope this work inspires readers and points ways forward towards addressing such difficult questions. 

\begin{acks}

We acknowledge the support of the Natural Sciences and Engineering Research Council of Canada (NSERC), RGPIN-2024-04049.

\end{acks}

\bibliographystyle{ACM-Reference-Format}
\bibliography{citations}

\appendix

\section{Unpacking ``formality''} \label{appendix-formality}

Although not necessary for understanding our main text, we have found it useful to critically reflect on the social construction of formality and terms like ``formal.'' This helps think through how established notations are social accomplishments, beyond technical ones. 

Today, ``formal'' appears appended to terms like ``languages,'' ``models,'' ``grammar,'' ``proof,'' ``system'', and ``semantics,'' across disciplines such as linguistics, computer science, logic, and mathematics. Typically in this usage a formal system is defined in notation borrowed from mathematical logic, with separate sections for syntax (representation) and semantics (meaning). This sense of formal appears to emerge from Frege and later Hilbert's formalist program~\cite{defining_formality}. Semantics in programming languages can be defined operationally (as rules on how patterns of notation are transformed), axiomatically or denotationally. In design, Shipman \& Marshall define ``formalisms'' as ``abstract representations that describe and constrain some aspect of their work or its content,'' requiring ``formalization'' of user ideas---users must learn a formal notation and encode their ideas into it, in order to gain the power of the machine. Interfaces risk ``imposing... too great a level of formality from their users,'' which can be ``harmful'' \cite[p. 334]{shipman1999formality}.

However, before and apart from its usage in computing, ``formal'' has a long history with other meanings. To understand this, we clustered definitions and etymological usage of the words ``formality,'' ``formal,'' ``informal'', and ``formalization'' on Oxford English Dictionary, Merriam-Webster, and Etymonline,\footnote{For clarity of presentation, various definitions from these source are quoted in this section without in-line citation. Definitions of words can also reference one another; for instance, formalization is defined as the process of "making formal": ``To give formal or definite shape to; to give formal being to'' \cite{oxfordenglishdictionary}.} to compile the following list:

\begin{enumerate}
    \item[F1] The formal are rules and procedures to follow (e.g., laws, rituals, ceremonies).
    \item[F2] The formal is precise, explicit, and unambiguous.
    \item[F3] The formal is mutually established by a community (cultural, social, disciplinary, etc).
    \item[F4] The formal conveys power and is defined by the powerful.
    \item[F5] The formal is rigid and strict, even ``exceedingly'' so.
    \item[F6] The formal is the domain of logic and reason.
    \item[F7] The formal as \textit{de-semantification}:  attention to representation and syntax to the extent of disregard for underlying meaning, i.e., ``abstraction from all meaning whatsoever''~\cite[p. 318]{defining_formality}
\end{enumerate}

``Formal'' emerged around 1390 AD in the context of ``established'' ceremonies, rituals, and laws---e.g., ``according to recognized forms, or to the rules of art or law.''\footnote{\url{https://www.oed.com/dictionary/formal_adj?tab=meaning_and_use\#3806070}} %
\textit{Formality} emerged later around the 1600s and is defined as "to make formal" and "conformity to established rule" (OED) with allusions to law. %
These early usages underscore that the formal is not primarily a technical thing, but a social one: what is considered formal is socially decided upon based on ``established rule.''  %
The term ``informal'' emerged around the mid-1500s to mean "not done or made according to a recognized or prescribed form; not observing established procedures or rules; unofficial; irregular,'' "without ceremony", "casual and relaxed" \cite{oxfordenglishdictionary}. The usage of ``established rule'' and ``ceremony'' implies that the quality of formality of a new system cannot be decided by the average individual, only by a community.

Another aspect worth reflection is how the formal appears in tension with content (F7). While the formal does hold connotations of "precision" and systems of rules that are established and adhered to, it also contains references to how this precision becomes "excessively rigid" or "mere ceremony" and de-emphasizes meaning in favor of form---\textit{the formal risks the death of meaning.} By contrast, the informal is "without ceremony" (notice the use of the same term---ceremony) and emphasizes content over form: the informal utterance holds functional, rather than performative value. A formal\textit{ism} has a "corresponding de-emphasis of content" in favor of form---focusing on representation over concepts---and is described as "rigid or merely conventional observance of forms," ``strict or excessive adherence to prescribed forms,'' "excessive regularity" or "stiffness" to the extent of even sometimes being "required to be done for form's sake." %
English idioms like ``it's only a formality'' reflect this double-meaning. In the philosophy of mathematics, \textit{formalism} is a perspective that believes that game-like manipulation of symbolic notation in proofs is all there is: ``mathematical utterances have no meaning... rather mathematics is a calculus in which `empty' symbol strings are transformed according to fixed rules'' \cite{weir2011formalism}. However, this notion of formality---as a formal grammar like Backus-Naur form---still has informalities in how it is interpreted. For instance, a ``formal'' grammar like Backus-Naur form has ambiguities in the order to parse an expression, ambiguities in how to symbols key to the syntax, like \texttt{::=}, may be used in expressions. No formal system can be both complete and consistent~\cite{raattkainen2005philosophical}. %

\subsubsection*{\bf The square of formality} Our etymological analysis calls our attention to the social construction of formality. We might conclude that formalization can proceed in one of two ways: %

The first, more common way is that an informal, ambiguous abstraction is given concrete shape and meaning through a process of \textit{translation} to a \textit{pre-existing} notation that is \textit{already} recognized within a community (e.g., prompting an LLM for Python code).  %

The second, more mysterious way formalization is achieved is that a new notation undergoes a cognitive, social, and material process whereby it is honed and refined and emerges with a stamp of "formal" approval: a \textit{new} notational system emerges. %

We might remember the difference by a `square of formality' (Figure~\ref{fig:square-of-formality}) revealing how formalization is a triangulation of translation of an idea into a established target notation (horizontal movement) as well as the active social construction of a new forms (vertical movement). %
Unlike horizontal movement, vertical movement does not have the luxury of piggybacking on a pre-existing system, but rather must ``earn'' its right to be considered formal. %

These two ``formalizing movements'' are not completely distinct. We have teased them out for theoretical power, but nowadays---surrounded as we are by notations---we generally have both horizontal and vertical movement when creating a new notation. %
We might say that the circumstances determine the degree of `vertical' movement required: many new notations seem spurred on by revolutionary developments in science, specifically by the need to tame new concepts~%
\cite{latour2011drawing}. Although this abstraction is a simplification, it may prove useful as a `tool to think with.' %

\section{Notations Considered in Comparative Analysis} \label{appendix-notations}

In our historical review, we considered: sheet music notation, dance notations like Labanotation and DanceWriting, Feynman-Dyson diagrams, Penrose diagrams, categorical quantum mechanics, quantum circuits, early programming notations, the Python and JavaScript languages and their communities, Markdown, micro-gesture notation~\cite{micro-gesture-notation-2023}, chemical notations from Berzelius' molecular formula to later structural formula by Lewis, Kekul\'{e}, and others, written sign language notations like SignWriting and Stokoe notation, UML diagrams, ``nerve net'' diagrams in early computing~\cite{kleene1956representation, mcculloch1943logical}, mathematical notations from Cajori's classic review~\cite{cajori1993history}, juggling notations, and Playfair's data visualizations. Table~\ref{tab:notation-survey} shows this information alongside key data sources we used, both primary documents and published historical accounts. Note that not all popular notations have complete histories, with details missing especially around inventor's initial construction processes.

\begin{table*}[t]
\centering
\small
\renewcommand{\arraystretch}{1.2}
\arrayrulecolor{black}  %
\begin{tabular}{p{3.2cm}|p{4cm}|p{3cm}|p{4cm}}
\toprule
\textbf{Name of Notation} & \textbf{Inventor(s)} & \textbf{Secondary Source(s)} & \textbf{Primary Source(s)} \\
\midrule

Sheet music notation & Various since antiquity & Goodman~\cite{goodman1976languages}, Ingold~\cite{ingold2016lines}, research in ethnomusicology \cite{schuiling2019notationcultures, lilliestam1996playing} & \\ \hline

Feynman--Dyson diagrams & Richard Feynman, Freeman Dyson, others in the decades since & Kaiser \cite{kaiser2019drawing}; \cite{meynell2008feynman, brown2018feynman} & \cite{dyson1949radiation, feynman1949space} \\ \hline

Penrose diagrams & Roger Penrose & Wright \cite{wright2013origins, wright2012advantages} & \cite{penrose1996space} \\ \hline

String diagrams in quantum mechanics & Bob Coecke (categorical quantum mechanics), Roger Penrose (tensor notation) & & Coecke \cite{coecke2017picturing}, Penrose \cite{penrose1971applications} \\ \hline

Quantum circuits & Various (e.g., Feynman, Barenco et al.) & & Toffoli \cite{toffoli1981bicontinuous}, Feynman \cite{feynman1986quantum}, Barenco et al.~\cite{barenco1995elementary} \\ \hline

Labanotation & Rudolf Laban, later built upon and popularized by Ann Hutchinson Guest & \cite{mccaw2023artlaban, mccaw2012laban, maletic1987bodylaban, knust1959introduction} & Some original Laban works republished in \cite{mccaw2023artlaban}; \cite{guest2013labanotation} \\ \hline

SignWriting \& DanceWriting & Valerie Sutton & \cite{bianchini2012writing} & Sutton's biographic histories; official videos on the SignWriting Channel \cite{sutton1973sutton, sutton2015history, signwriting2014, DeafPerspectivesOnSignWriting} \\ \hline

Early programming notations & Konrad Zuse, John von Neumann, Herman Goldstine, Grace Hopper, John Backus, early FORTRAN and ALGOL groups, many others & \cite{arawjo2020write, history_algol, knuth1980early, bemer1971view} & \cite{naur1968successes, backus1978history, backus1957fortran, goldstine1947planning, iverson2007notation, zuse1993computer} \\ \hline

Python & Guido van Rossum & & Python documentation and PEPs (e.g., \cite{pep1, pep13, pep257, pep572}); oral history~\cite{venners2003making, WhyPythonWasCreated}; van Rossum USENET posts around 1991; van Rossum's ``Transfer of Power'' announcement \cite{vanrossum2018transfer_of_power} \\ \hline

JavaScript ES6 & Brendan Eich, others & & ECMAScript standards; TC39 proposals; Eich's blog \cite{eich2011jsconf, eich2011harmony} \\ \hline

Markdown and CommonMark & John Gruber, Aaron Swartz; CommonMark team including John MacFarlane, Martin Woodward, Jeff Atwood & & \cite{commonmark, atwood2014commonmark, Gruber2004Markdown} \\ \hline

Micro-gesture notation & Adrien Chaffangeon Caillet, Alix Goguey, and Laurence Nigay & & \cite{micro-gesture-notation-2023} \\ \hline

Chemical molecular formulas and later structural formulas & Jöns Jakob Berzelius, G. N. Lewis, August Kekul\'{e}, others & Klein \cite{klein2003experiments}, \cite{wiswesser1985historic, hein1966kekule} & \cite{berzelius1814attempt, berzelius1833betrachtungen, lewis1916atom, kekule1858ueber, kekule1865constitution} \\ \hline

Stokoe notation & William Stokoe & \cite{mccarty2004signstokoe} & \cite{stokoe2005sign} \\ \hline

UML diagrams & Booch, Rumbaugh, and Jacobson (the ``Three Amigos'') & \cite{platt2015evolution, visualparadigm} & \cite{booch1981describing} \\ \hline

“Nerve net’’ diagrams & McCulloch and Pitts; Kleene & & \cite{kleene1956representation, mcculloch1943logical} \\ \hline

Mathematical notations & Various since antiquity & Cajori \cite{cajori1993history} & \\ \hline

Juggling notations & Claude Shannon, others, mainly in hobbyist contexts & \cite{lewbel1994juggling} & \cite{simpson1986juggling, beek1995sciencejuggling, tiemann1991notationjuggling, walker1982variationsjugglers} \\ \hline

Line and scatter plots & William Playfair (arguably) & \cite{spence2017william} & \cite{playfair1786commercial} \\

\bottomrule
\end{tabular}
\caption{Survey of notations considered, their inventor(s), and consulted primary and secondary sources.}
\label{tab:notation-survey}
\end{table*}

\end{document}